\documentclass{tlp}

\usepackage[utf8]{inputenc}
\usepackage[english]{babel}

\usepackage{amsmath}
\usepackage{amssymb}
\usepackage{stmaryrd}
\usepackage{latexsym}
\usepackage{algorithm}
\usepackage{algpseudocode}
\usepackage{marvosym}

\newtheorem{definition}{Definition}
  [section]
\newtheorem{example}{Example}
  [section]
\newtheorem{lemma}{Lemma}
  [section]
\newtheorem{theorem}{Theorem}
  [section]

\usepackage[T1]{fontenc}
\usepackage[scaled=.81]{beramono}
\usepackage{bm}

\usepackage{multirow}
\usepackage{subcaption}
\usepackage{graphicx}
\usepackage{booktabs}

\usepackage{xcolor}
\usepackage{listings}
\definecolor{ciaoframe}{rgb}
  {0,0,0.3}
\definecolor{ciaostring}{rgb}
  {0.6,0.46,0.33}
\definecolor{ciaooperators}{rgb}
  {0.1,0.15,0.6}
\definecolor{ciaokeywords}{rgb}
  {0.1,0.15,0.6}
\definecolor{ciaoassertions}{rgb}
  {0.1,0.15,0.6}
\definecolor{ciaotrust}{rgb}
  {200,130,0}
\definecolor{ciaocheck}{rgb}
  {0.1,0.2,0.8}
\definecolor{ciaochecked}{rgb}
  {0.2,0.34,0.1}
\definecolor{ciaotrue}{rgb}
  {0.2,0.34,0.1}
\definecolor{ciaofalse}{rgb}
  {0.6,0.0,0.09}
\definecolor{ciaoprops}{rgb}
  {0.1,0.2,0.8}
\definecolor{ciaocomment}{rgb}
  {0.6,0.0,0.09}
\lstdefinelanguage{Ciao}{%
  language=Prolog,
  xleftmargin=4mm,
  breaklines=true,
  tabsize=2,
  breaklines=true,breakatwhitespace=true,
  basicstyle=\rm\ttfamily,
  showlines=true,
  showspaces=false,
  showtabs=false,
  mathescape=true,
  escapeinside={(*}{*)},
  commentstyle=\color{ciaocomment},
  stringstyle=\color{ciaostring},
  showstringspaces=false,
  deletekeywords={true,pp},
  keywordstyle={\color{ciaooperators}\bfseries},
  classoffset=1,
  otherkeywords={@,>,<,>=,=<,.,;,-,!,=,*,\&,+,:-,[,],|,->,:,:=,\#,?},
  keywordstyle={\color{ciaokeywords}\bfseries},
  classoffset=2,
  morekeywords={module,use_module,dynamic,export,import,impl_defined,trait,impl},
  keywordstyle={\color{ciaokeywords}\bfseries},
  morekeywords={pred,prop,calls,success,comp,compat,inst,modedef,regtype},
  keywordstyle={\color{ciaoassertions}\bfseries},
  classoffset=4,
  morekeywords={trust,trust_default,entry},
  keywordstyle={\color{ciaotrust}\bfseries},
  classoffset=5,
  morekeywords={check},
  keywordstyle={\color{ciaocheck}\bfseries},
  classoffset=6,
  morekeywords={checked},
  keywordstyle={\color{ciaochecked}\bfseries},
  classoffset=7,
  morekeywords={true},
  keywordstyle={\color{ciaotrue}\bfseries},
  classoffset=8,
  morekeywords={false},
  keywordstyle={\color{ciaofalse}\bfseries},
  classoffset=9,
  morekeywords={
    nat,int,nnegint,flt,atm,term,num,var,list,ground,mshare,
    rsize,cardinality,not_fails,fails,exp,cost,costb,steps_ub,steps_lb,
    size_ub,size_lb,covered,mut_exclusive,head_cost,literal_cost,
    is_det,det,nondet,semidet,multi,terminates,steps_o,resource,socket,
    seff,string,not_further_inst,nonground,unknown,unreachable,
    sorted_int_list, sorted, no_choicepoints, shfr, any, builtin,
    prefix, int_op, pp, p_nat, p_neg, p_nat_nat, dutch_cmp, wb, rwb,
    rwbo, rt1, rt2, lg, lge, lgLGe, negz, neg, zero, num_cmp, req, res,
    handler, t, t_cmp, t_sort
  },
  keywordstyle={\color{ciaoprops}\bfseries},
  classoffset=0
}
\lstdefinelanguage{CiaoLong}{%
  language=Ciao,
  basicstyle=\small\rm\ttfamily,
  frame=l,
  rulecolor=\color{ciaoframe},
  numbers=left,
  numberstyle=\tiny,
  numbersep=6pt,
  xleftmargin=4pt,xrightmargin=4pt
}
\newcommand{\ciaoinline}[1]
  {{\rm\lstinline[language=Ciao]{#1}}}
\usepackage[normalem]{ulem}
\usepackage{xspace}

\usepackage[%
  colorinlistoftodos,
  textsize=scriptsize,
  textwidth=1.3cm
]{todonotes}

\usepackage{url}
\usepackage{hyperref}
\hypersetup{%
  colorlinks=true,
  citecolor=blue,
  linkcolor=blue,
  urlcolor=blue,
  filecolor=blue,
  pdftitle={Hiord\#: An Approach to the Specification and Verification
    of Higher-Order (C)LP Programs},
  pdfauthor={Marco Ciccal\`{e}, Daniel Jurjo-Rivas, Jose F.~Morales,
    Pedro L\'{o}pez-Garc\'{i}a, Manuel V.~Hermenegildo},
  pdfsubject={Computer Science},
  pdfkeywords={Higher-Order, Static Analysis, Assertions, Abstract
    Interpretation, (Constraint) Logic Programming},
  bookmarksopen=true,
  bookmarksnumbered=true
}
\usepackage[capitalize,noabbrev]{cleveref}
\crefformat{section}{\S#2#1#3}
\Crefformat{section}{\S#2#1#3}
\usepackage{ifthen}
\newboolean{final}\setboolean{final}{true}

\newcommand{\hiord}
  {Hiord$\, ^\sharp$}

\renewcommand{\implies}
  {\ensuremath{\Rightarrow}}
\renewcommand{\iff}
  {\ensuremath{\Leftrightarrow}}
\renewcommand{\emptyset}
  {\varnothing}
\newcommand{\pholder}
  {\_}
\newcommand{\setcomp}[2]
  {\ensuremath{\{#1~|~#2\}}}
\newcommand*{\seq}[2][n]
  {{#2_{1},\allowbreak\ldots,\allowbreak #2_{#1}}}
\newcommand{\tuple}[1]
  {\langle #1 \rangle}
\newcommand{\envend}
  {\hfill$\Box$}
\newcommand{\quantSep}
  {\ensuremath{\ldotp\;}}

\newcommand{\clp}
  {(C)LP\xspace}
\newcommand{\plai}
  {\textsf{PLAI}\xspace}
\newcommand{\ciao}
  {\textsf{Ciao}\xspace}
\newcommand{\ciaopp}
  {\textsf{CiaoPP}\xspace}
\newcommand{\ciaochecked}
  {\texttt{\bfseries\color{ciaochecked}checked}\xspace}
\newcommand{\ciaofalse}
  {\texttt{\bfseries\color{ciaofalse}false}\xspace}
\newcommand{\ciaocheck}
  {\texttt{\bfseries\color{ciaocheck}check}\xspace}

\newcommand{\prog}
  {\ensuremath{P}}
\newcommand{\entails}
  {\models}
\newcommand{\defn}[1]
  {\ensuremath{\mathsf{defn}(#1)}}
\newcommand{\ar}[1]
  {\ensuremath{\mathsf{ar}(#1)}}
\newcommand{\pred}[2]
  {\ensuremath{#1~\texttt{:-}~#2}}
\newcommand{\card}[1]
  {\ensuremath{|#1|}}

\newcommand{\emptyGoal}
  {\square}
\newcommand{\cquery}
  {\ensuremath{Q}}
\newcommand{\Q}
  {\ensuremath{\mathcal{Q}}}
\newcommand{\cqueries}
  {\ensuremath{\Q}}
\newcommand{\state}[2]
  {\langle #1\, |\, #2 \rangle}
\newcommand{\red}[2]
  {#1\leadsto #2}
\newcommand{\redstar}[2]
  {#1\leadsto^* #2}
\newcommand{\derivs}[2]
  {\ensuremath{\mathsf{derivs}(#1,#2)}}
\newcommand{\answers}[2]
  {\ensuremath{\mathsf{answers}(#1,#2)}}

\newcommand{\cdom}
  {\ensuremath{D}}
\newcommand{\adom}
  {\ensuremath{D^\sharp}}
\newcommand{\aquery}
  {\ensuremath{Q^\sharp}}
\newcommand{\aqueries}
  {\ensuremath{\Q^\sharp}}
\newcommand{\absint}[1][\prog]
  {\ensuremath{\llbracket #1 \rrbracket_{\aqueries}^\sharp}}
\newcommand{\acheck}[2]
  {\ensuremath{\mathsf{acheck}(#1,#2)}}
\newcommand{\atriple}[1][]
  {\ensuremath{\tuple{L_{#1},\lambda^c_{#1},\lambda^s_{#1}}}}

\newcommand{\trivSucc}[2][\theta]
  {\ensuremath{#1\Rightarrow_{\prog}#2}}
\newcommand{\notTrivSucc}[2][\theta]
  {\ensuremath{#1\not\Rightarrow_{\prog}#2}}
\newcommand{\trivset}[1]
  {\ensuremath{#1^{\natural}}}
\newcommand{\abstrivsubset}[2]
  {\ensuremath{#1^{\sharp-}}}
\newcommand{\abstrivsupset}[2]
  {\ensuremath{#1^{\sharp+}}}

\newcommand{\pre}
  {\ensuremath{\mathit{Pre}}}
\newcommand{\post}
  {\ensuremath{\mathit{Post}}}
\newcommand{\acCall}[2]
  {\ensuremath{\mathsf{calls}(#1,#2)}}
\newcommand{\acSucc}[3]
  {\ensuremath{\mathsf{success}(#1,#2,#3)}}
\newcommand{\aid}[1][]
  {\ensuremath{\ell_{#1}}}

\newcommand{\A}
  {\ensuremath{\mathcal{A}}}
\newcommand{\falseAc}
  {\ensuremath{\mathcal{E}}}
\newcommand{\checklit}[2]
  {\ensuremath{\mathsf{check}(#1,#2)}}
\newcommand{\exstate}[3]
  {\langle #1\, |\, #2\, |\, #3 \rangle}
\newcommand{\redA}[2]
  {#1\leadsto_{\A} #2}
\newcommand{\redstarA}[2]
  {#1\leadsto^*_{\A} #2}
\newcommand{\derivsA}[3][\A]
  {\ensuremath{\mathsf{derivs}_{#1}(#2,#3)}}

\newcommand{\anonyAssrt}
  {\ensuremath{{}^{\circ}\hspace{-0.4em}A}}
\newcommand{\anonyACond}
  {\ensuremath{{}^{\circ}\hspace{-0.15em}C}}
\newcommand{\anonyPre}
  {\ensuremath{{}^{\circ}\hspace{-0.25em}\mathit{Pre}}}
\newcommand{\anonyPreBold}
  {\ensuremath{{}^{\circ}\hspace{-0.25em}\bm{\mathit{Pre}}}}
\newcommand{\anonyPost}
  {\ensuremath{{}^{\circ}\hspace{-0.25em}\mathit{Post}}}
\newcommand{\anonyPostBold}
  {\ensuremath{{}^{\circ}\hspace{-0.25em}\bm{\mathit{Post}}}}
\newcommand*{\instAnonyAssrt}[2][]
  {\ensuremath{\anonyAssrt_{#1}\hspace{-0.1em}|_{#2}}}
\newcommand{\instPredprop}[1]
  {\ensuremath{\Pi|_{#1}}}
\newcommand{\p}
  {\ensuremath{p(\bar{v})}}
\newcommand{\anonyHead}
  {\ensuremath{\pholder(\bar{v})}}
\newcommand{\conf}
  {\ensuremath{\mathbin{\prec}}}
\newcommand{\nconf}
  {\ensuremath{\mathbin{\nprec}}}
\newcommand{\aconf}
  {\ensuremath{\mathbin{\prec^{\sharp-}}}}
\newcommand{\aconfover}
  {\ensuremath{\mathbin{\prec^{\sharp+}}}}
\newcommand{\naconf}
  {\ensuremath{\mathbin{\nprec^{\sharp+}}}}

\begin{document}
\lefttitle{M.~Ciccal{\`{e}} et al.}
\jnlPage{1}{19}
\jnlDoiYr{2025}
\doival{\href{https://doi.org/10.1017/S147106842510015X}{10.1017/S147106842510015X}}
\title[\hiord]
  {\vspace{2em}\hiord: An Approach to the Specification and\\
     Verification of Higher-Order (C)LP Programs%
     \thanks{Partially funded by MICIU
       projects
       CEX2024-001471-M \emph{María de Maeztu}
       and TED2021-132464B-I00 \emph{PRODIGY}, as well as by the
       Tezos foundation.
       We would also like to thank the anonymous reviewers for their
       very useful and constructive feedback.}}
\begin{authgrp}
  \author{%
    \href{https://orcid.org/0009-0000-8821-0587}{MARCO CICCALÈ},
    \href{https://orcid.org/0000-0001-6215-1080}{DANIEL JURJO-RIVAS} and
    \href{https://orcid.org/0000-0001-9782-8135}{JOSE F.~MORALES}\\[.6mm]}
  \affiliation{%
    Universidad Polit\'{e}cnica de Madrid (UPM),
    IMDEA Software Institute, Madrid, Spain\\[.6mm]
    {\rm(}e-mails:
    {\fontfamily{cmtt}\selectfont\upshape\href{mailto:m.ciccale@alumnos.upm.es}{m.ciccale@alumnos.upm.es}{\rm\it, }%
       \href{mailto:marco.ciccale@imdea.org}{marco.ciccale@imdea.org}{\rm\it, }%
       \href{mailto:daniel.jurjo@alumnos.upm.es}{daniel.jurjo@alumnos.upm.es}{\rm\it, }%
       \href{mailto:daniel.jurjo@imdea.org}{daniel.jurjo@imdea.org}{\rm\it, }%
       \href{mailto:josefrancisco.morales@upm.es}{josefrancisco.morales@upm.es}{\rm\it, }%
       \href{mailto:josef.morales@imdea.org}{josef.morales@imdea.org}}%
     {\rm)}\\[1mm]}
   \author{%
     \href{https://orcid.org/0000-0002-1092-2071}{PEDRO LÓPEZ-GARCÍA}\\[.6mm]}
   \affiliation{%
     Spanish Council for Scientific Research,
     IMDEA Software Institute, Madrid, Spain\\[.6mm]
    {\rm(}e-mails:
    {\fontfamily{cmtt}\selectfont\upshape\href{mailto:pedro.lopez@csic.es}{pedro.lopez@csic.es}{\rm\it, }%
       \href{mailto:pedro.lopez@imdea.org}{pedro.lopez@imdea.org}}%
     {\rm)}\\[1mm]}
   \author{%
     \href{https://orcid.org/0000-0002-7583-323X}{MANUEL V.~HERMENEGILDO}\\[.6mm]}
   \affiliation{%
     Universidad Polit\'{e}cnica de Madrid (UPM),
     IMDEA Software Institute, Madrid, Spain\\[.6mm]
    {\rm(}e-mails:
    {\fontfamily{cmtt}\selectfont\upshape\href{mailto:manuel.hermenegildo@upm.es}{manuel.hermenegildo@upm.es}{\rm\it, }%
       \href{mailto:manuel.hermenegildo@imdea.org}{manuel.hermenegildo@imdea.org}}%
     {\rm)}\\[-4mm]}
\end{authgrp}
\history{\sub{22 July 2025;} \rev{22 July 2025;} \acc{27 July 2025}}
\maketitle
\begin{abstract}
  Higher-order constructs enable more expressive and concise code by
  allowing procedures to be parameterized by other procedures.
  Assertions allow expressing partial program specifications, which
  can be verified either at compile time (statically) or run time
  (dynamically).
  In higher-order programs, assertions can also describe higher-order
  arguments.
  While in the context of (constraint) logic programming ((C)LP),
  run-time verification of higher-order assertions has received some
  attention, compile-time verification remains relatively unexplored.
  We propose a novel approach for statically verifying higher-order
  (C)LP programs with higher-order assertions.
  Although we use the \ciao assertion language for illustration, our
  approach is quite general and we believe is applicable to similar
  contexts.
  Higher-order arguments are described using predicate properties -- a
  special kind of property which exploits the (\ciao) assertion
  language.
  We refine the syntax and semantics of these properties and introduce
  an abstract criterion to determine conformance to a predicate
  property at compile time, based on a semantic order relation
  comparing the predicate property with the predicate assertions.
  We then show how to handle these properties using an abstract
  interpretation-based static analyzer for programs with first-order
  assertions by reducing predicate properties to first-order
  properties.
  Finally, we report on a prototype implementation and evaluate it
  through various examples within the \ciao system.
\end{abstract}
\section{Introduction}
\label{sec:Introduction}
Abstraction is a fundamental principle in computer science often used
for managing complexity.
Higher-order constructs are a form of abstraction that enables writing
code that is more concise and expressive by allowing procedures to be
parameterized by other procedures, resulting in more modular and
maintainable code.
(Constraint) logic programming languages like
\textsf{Prolog}~\citep{prolog50-tplp-short} and functional programming
languages like \textsf{Haskell}~\citep{haskell-standard-2010} have
included different forms of higher-order since their early days, and
languages from other programming paradigms like \textsf{Java} or
\textsf{C++} have adopted them later on.
In particular, \textsf{Prolog} systems allow defining higher-order
predicates and making higher-order calls.
For example, the query:
\ciaoinline{?-$~$filter(even,[7,4,9],L)},
passes the term \ciaoinline{even} as an argument to the higher-order
predicate \texttt{filter/3}, which \emph{applies} the
\ciaoinline{even/1} predicate to each element of the input list,
selecting those that succeed, yielding \ciaoinline{L$~$=$~$[4]}.
Assertions are linguistic constructs for writing partial program
specifications, which can then be verified or used to detect
deviations in program behavior w.r.t.~such specifications.
The \emph{assertion-based approach} to program
verification~\citep{prog-glob-an-short,assrt-theoret-framework-lopstr99-short,verifly-2021-tplp-short}
differs from other approaches such as strong type
systems~\citep{DBLP:conf/ifip2/Cardelli89-short} in that assertions
are optional and can include properties that are undecidable at
compile time, and thus some checking may need to be relegated to run
time.
Hence, the assertion-based approach is closer to gradual typing in
functional languages~\citep{Siek06gradualtyping}.
The combination of higher-order predicates and assertions in the \clp
context was already explored by~\cite{asrHO-ppdp2014-shorter}.
This work introduced the notion of \emph{predicate properties}, a
special kind of properties that allow using the full power of the
(\ciao) assertion language for describing the higher-order arguments
of procedures.
This work also proposed an operational semantics for
\emph{dynamically} checking higher-order \clp programs annotated with
such higher-order assertions.
However, the \emph{static} verification of programs with higher-order
assertions was not addressed in that work, and remains relatively
unexplored since other related work in \clp that supports higher
order~(\emph{e.g.}, \citet{Miller,goedel,mercury-jlp}) generally
adheres to the strong typing model.
In this work we propose a novel approach for the compile-time
verification of higher-order \clp programs with assertions describing
higher-order arguments.
We present a refinement of both the syntax and the semantics of
predicate properties (\cref{sec:spec-predprops}).
Next, we define an abstract criterion to determine whether a predicate
conforms to a predicate property at compile time, based on a semantic
order relation between the definition of a predicate property and the
partial specification of a predicate (\cref{sec:predprop_conf}).
Then, we introduce an approach for ``casting'' predicate usage in a
program analysis-friendly manner that enhances and complements the
proposed abstract criterion (\cref{sec:wrappers}).
We also propose a technique for dealing with these properties using an
abstract interpretation-based static analyzer for programs with
first-order assertions, by representing predicate properties as
first-order properties that are natively understood by such an
analyzer (\cref{sec:ground-predprops}).
Finally, we present a prototype implementation of these techniques and
study its application to a number of examples (\cref{sec:exp}).
For concreteness, we use in our presentation the
\ciao~\citep{ciao-design-tplp-shorter} assertion language, and make
use of its \ciaopp
preprocessor~\citep{ciaopp-sas03-journal-scp-short}, that combines
both static and dynamic analysis.  However, we believe the approach is
quite general and flexible, and can be applied, at least conceptually,
to similar gradual approaches.
\section{Preliminaries and notation}
\label{sec:prel}
Variables start with a capital letter.
The set of terms is inductively defined as follows: (1) variables are
terms (2) if $f$ is an $n$-ary function symbol and $\seq{t}$ are
terms, then $f(\seq{t})$ is a term.
We use the overbar notation $(\bar{\cdot})$ to denote a finite
sequence of elements (\emph{e.g.}, $\bar{t} \equiv \seq{t}$), and write
$\card{(\bar{\cdot})}$ for representing its length.
An \emph{atom} has the form $p(\bar{t})$ where $p$ is an $n$-ary
predicate symbol, and $\bar{t}$ are terms.
The function $\ar{p}$ denotes the arity of a predicate $p$.
A \emph{higher-order atom} has the form $X(\bar{t})$ where $X$ is a
variable and $\bar{t}$ are terms.
(Note that variables are \emph{not} allowed in the function symbol
position of terms, only in literals).
A \emph{constraint} is a conjunction of expressions built from
predefined predicates whose arguments are constructed using predefined
functions and variables, for example, $X-Y > \mathit{abs}(Z)$.
A \emph{literal} is either an atom, a higher-order atom, or a
constraint.
\emph{Negation} is encoded as finite failure, supported through a program
expansion.
A \emph{goal} is a finite sequence of literals.
A \emph{rule} has the form $\pred{H}{B}$ where $H$, the \emph{head},
is an atom and $B$, the \emph{body}, is a possibly empty finite
sequence of literals.
A \emph{higher-order constraint logic program}, or \emph{higher-order
  program} $\prog$ is a finite set of rules.
We use $\sigma$ to represent a variable renaming, and $\sigma(L)$ or
$L\sigma$ to represent the result of applying $\sigma$ to a syntactic
object $L$.
The \emph{definition} of an atom $L$ in a program, $\defn{L}$, is the
set of renamed program rules s.t.~each renamed rule has $L$ as its
head.
We assume that all rule heads are \emph{normalized}, that is, $H$ is an
atom of the form $\p$ where $\bar{v}$ are distinct variables.
Let $\bar{\exists}_L \theta$ denote the projection of the constraint
$\theta$ onto the variables of $L$.
We denote \emph{constraint entailment} by $\theta_1\entails \theta_2$.
\subsection{Operational semantics of higher-order programs}
The operational semantics of a higher-order program $\prog$ is given
in terms of its \emph{derivations}, which are sequences of
\emph{reductions} between \emph{states}.
A state $\state{G}{\theta}$ consists of a goal $G$, and a constraint
store (or store) $\theta$.
We denote sequence concatenation by (::).
We assume for simplicity that the underlying constraint solver is
complete and projection exists.
We use $\red{S}{S'}$ to indicate that a reduction step can be applied
to state $S$ to obtain state $S'$.
Naturally, $\redstar{S}{S'}$ indicates that there is a sequence of
reduction steps from $S$ to $S'$.
A state $S = \state{L::G}{\theta}$ where $L$ is a literal, is
\emph{reduced} to a state $S'$ as follows:
\begin{enumerate}
\item If $L$ is a constraint and $\theta\land L$ is satisfiable, then
  $S' = \state{G}{\theta \land L}$.
\item If $L$ is an atom and $\exists(\pred{L}{B})\in\defn{L}$, then
  $S' = \state{B::G}{\theta}$.
\item If $L$ is a higher-order atom
  of the form $X(\bar{t})$, then
  $S' = \state{p(\bar{t}) :: G}{\theta}$ given that
  $\exists p \in \prog \quantSep \theta \entails (X = p) \wedge
  \ar{p}=\card{\bar{t}}$.
\end{enumerate}

Let $L$ be an atom, $S=\state{L::G}{\theta}$, $S'=\state{G}{\theta'}$,
and suppose $\redstar{S}{S'}$.
We refer to $S$ as a \emph{call state} for $L$, and $S'$ as a
\emph{success state} for $L$.
A \emph{query} $\cquery$ is a pair $(L,\theta)$, where $L$ is a
literal and $\theta$ a store for which the (C)LP system starts a
computation from state $\state{L}{\theta}$.
The set of all derivations of $\prog$ from a query $\cquery$ is
denoted $\derivs{\prog}{\cquery}$, and this notation is naturally
extended to a set of queries $\cqueries$.
Let $D_{[-1]}$ denote the last state of any derivation $D$.
A finite derivation from a query $\cquery$ is \emph{finished} if
the last state in the derivation cannot be reduced.
A finished derivation from a query $\cquery$ is \emph{successful}
if the last state is of the form $\state{\emptyGoal}{\theta'}$, where
$\emptyGoal$ denotes the empty goal sequence.
In that case, the constraint $\bar{\exists}_{L}\theta'$ is an
\emph{answer} to $\cquery$.
We denote by $\answers{\prog}{\cquery}$ the set of answers of $\prog$
to a query $\cquery$.
A finished derivation is \emph{failed} if the last state is not of the
form $\state{\emptyGoal}{\theta}$.
A query $\cquery$ \emph{finitely fails} if all derivations in
$\derivs{\prog}{\cquery}$ are finished and have failed.
\subsection{Property formulas}
Conditions on the constraint store are stated as \emph{property formulas}.
A property formula is a DNF formula of \emph{property literals}.
A property literal is a literal corresponding to a special kind of
predicates called \emph{properties}.
Properties are typically defined in the source language, in the same
way as ordinary predicates but marked accordingly, and are required to
meet certain
conditions~\citep{prog-glob-an-short,assrt-theoret-framework-lopstr99-short}.
In particular, they are normally required to be checkable at run time
but not necessarily decidable at compile time, where they are safely
approximated.\footnote{\ciao assertions can also include \emph{global
    properties}, which may not always be checkable at run time (\emph{e.g.},
  termination), but we focus for brevity on the described types of
  assertions and properties.}
\begin{example}[Properties]
  \label{ex:props}
  The following program defines the properties \ciaoinline{list/1}
  (``being a list'') and \ciaoinline{prefix/2} (``being a prefix of a
  list''):

  \noindent\begin{minipage}{.37\textwidth}
\begin{lstlisting}
:- prop list/1. list([]).
list([_|Xs]) :- list(Xs).
\end{lstlisting}
  \end{minipage}
  \hfill
  \begin{minipage}{.62\textwidth}
\begin{lstlisting}[firstnumber=3]
:- prop prefix/2. prefix([],Ys) :- list(Ys).
prefix([X|Xs],[X|Ys]) :- prefix(Xs,Ys).
\end{lstlisting} 
  \end{minipage}
  The property formula \ciaoinline{(list(Xs),list(Ys),prefix(Xs,Ys))}
  states that \texttt{Xs} and \texttt{Ys} should be lists, and that
  \texttt{Xs} should be a prefix of \texttt{Ys}.
  This formula contains three property literals corresponding to the
  \ciaoinline{list/1} and \ciaoinline{prefix/2} properties.
\end{example}

We now recall an instrumental definition about properties
from~\citet{assrt-theoret-framework-lopstr99-short}:
\begin{definition}[Succeeds trivially]
  \label{def:succeeds-triv}
  A property literal $L$ \emph{succeeds trivially} for $\theta$ in a
  program $\prog$, denoted $\trivSucc{L}$, iff
  $\exists \theta'\in \answers{P}{(L,\theta)} \quantSep
  \theta\entails\theta'$.
  A property formula succeeds trivially for $\theta$ if all of the
  property literals of at least one conjunct of the formula succeeds
  trivially.
\end{definition}

Intuitively, a property literal (or formula) succeeds trivially if it
succeeds for $\theta$ in $\prog$ without adding new ``relevant''
constraints to $\theta$.
For example, \ciaoinline{list(X)} checks ``\texttt{X} being a list.''
\subsection{Traditional assertions}
Assertions are syntactic objects for expressing properties of programs
that must be satisfied at program execution.
We recall the herein relevant parts of the assertion schema
of~\citet{assert-lang-disciplbook-short}.
\emph{Traditional} (or \emph{first-order}) \emph{predicate} (or
\emph{pred}) assertions have the following syntax:
``\ciaoinline{:-$~$pred}$~\mathit{Pred}~$\ciaoinline{:}$~\pre~$\ciaoinline{=>}$~\post$\ciaoinline{.},''
where $\mathit{Pred}$ is a normalized atom representing a predicate,
and $\pre$ and $\post$ are property formulas.
They express that all calls to $\mathit{Pred}$ \emph{must} satisfy the
pre-condition $\mathit{Pre}$, and, if such calls succeed, the
post-condition $\mathit{Post}$ \emph{must} be satisfied.
If there are several \emph{pred} assertions, the $\pre$ field of at
least one of them must be satisfied.

\begin{example}[Assertions]
  \label{ex:take_assrt}
  The following assertions for the \texttt{take/3} predicate relating
  a list and its prefix:
\begin{lstlisting}
:- pred take(N,Xs,Ys) : (int(N), list(Xs)) => prefix(Ys,Xs).
:- pred take(N,Xs,Ys) : (list(Xs), prefix(Ys,Xs)) => int(N).
\end{lstlisting}
  restrict the meaning of \texttt{take/3} as follows:
\end{example}
%%% If I don't take the itemize out of the example env.,
%%% the margins get messed up?
\begin{itemize}\itshape
\item \ciaoinline{take(N,Xs,Ys)} must be called with \ciaoinline{Xs} bound to
  a list, and either \ciaoinline{N} bound to an integer or \ciaoinline{Ys}
  bound to a prefix of \ciaoinline{Xs}.
\item if \ciaoinline{take(N,Xs,Ys)} succeeds when called with \ciaoinline{N}
  bound to an integer and \ciaoinline{Xs} bound to a list, then
  \ciaoinline{Ys} must be bound to a prefix of \ciaoinline{Xs}.
\item if \ciaoinline{take(N,Xs,Ys)} succeeds when called with \ciaoinline{Xs}
  bound to a list and \ciaoinline{Ys} bound to a prefix of \ciaoinline{Xs},
  then \ciaoinline{N} must be bound to an integer.
\end{itemize}

We represent checks on the store by a set of assertions with a set of
\emph{assertion conditions}.
\begin{definition}[Assertion conditions]
  \label{def:ac}
  Given a predicate represented by a normalized atom $\mathit{Pred}$,
  and its corresponding set of assertions $\{\seq{A}\}$ with
  $A_i=\text{``\ciaoinline{:-$~$pred}$~\mathit{Pred}~$\ciaoinline{:}$~\mathit{Pre}_i~$\ciaoinline{=>}$~\mathit{Post}_i$\ciaoinline{.},''}$
  the set of \emph{assertion conditions} for $\mathit{Pred}$ is
  $\{C_0,\seq{C}\}$ with
  \[
    C_i =
    \begin{cases}
      \acCall{\mathit{Pred}}{\bigvee_{j=1}^n\, \mathit{Pre_j}} & i = 0\\
      \acSucc{\mathit{Pred}}{\mathit{Pre_i}}{\mathit{Post_i}} & i \in 1..n
    \end{cases}
  \]
  Condition $C_0$ encodes the checks that ensure that all calls to the
  predicate represented by $\mathit{Pred}$ are within those admissible
  by the set of assertions; we refer to it as the \emph{calls
    assertion condition}.
  Conditions $\seq{C}$ encode the checks for compliance of the
  successes for particular sets of calls, and we call them the
  \emph{success assertion conditions}.
\end{definition}

From this point on, we denote by $\A$ both the set of assertions of
the program and, interchangeably, its associated set of assertion
conditions.
Also, for a normalized atom $\mathit{Pred}$, $\A(\mathit{Pred})$
denotes only the assertions of $\A$ associated to the predicate
$\mathit{Pred}$.
\begin{example}[Assertion conditions]
  \label{ex:take_ac}
  The set of assertion conditions for the set of \emph{pred}
  assertions in~\cref{ex:take_assrt} is:
  \begin{align*}
    \mathsf{calls}&(%
      \text{\ciaoinline{take(N,Xs,Ys)}},~
        \text{\ciaoinline{(int(N), list(Xs))}}~
        \vee\,
        \text{\ciaoinline{(list(Xs), prefix(Ys,Xs))}})&\\[-1mm]
    \mathsf{success}&(%
      \text{\ciaoinline{take(N,Xs,Ys)}},~
      \text{\ciaoinline{(int(N), list(Xs))}},~
      \text{\ciaoinline{prefix(Ys,Xs)}})&\\[-1mm]
    \mathsf{success}&(%
      \text{\ciaoinline{take(N,Xs,Ys)}},~
      \text{\ciaoinline{(list(Xs), prefix(Ys,Xs))}},~
      \text{\ciaoinline{int(N)}})&
  \end{align*}
\end{example}
\subsection{Operational semantics of higher-order programs with
  traditional assertions}
This operational semantics checks whether assertion conditions hold or
not while computing the derivations from a query, halting the
derivation as soon as an assertion condition is violated.
For identifying a possible assertion condition violation, every
assertion condition $C$ is related to a unique label $\ell$ via a
mapping $\mathsf{label}(C) = \ell$.
States of derivations are now of the form
$\exstate{G}{\theta}{\falseAc}$, where $\falseAc$ denotes the set of
labels for falsified assertion conditions (with
$\card{\falseAc} \leqslant 1$); while such set is unnecessary if
execution halts upon an assertion condition violation, we include it
to keep the semantics presented in this paper close to that of
previous work.
A finished derivation from a query $\cquery=(L,\theta)$ is now
\emph{successful} if the last state is of the form
$\exstate{\emptyGoal}{\theta'}{\emptyset}$, \emph{failed} if the last
state is of the form $\exstate{L'}{\theta'}{\emptyset}$, and
\emph{erroneous} if the last state is of the form
$\exstate{L'}{\theta'}{\{\aid\}}$.
We also extend the set of literals with syntactic objects of the form
$\checklit{L}{\aid}$ where $L$ is a literal and $\aid$ is a label for
an assertion condition, which we call \emph{\textsf{check} literals}.
Thus, a \emph{literal} is now a constraint, an atom, a higher-order
atom, or a $\mathsf{check}$ literal.
We now recall the notion of \emph{Semantics with Assertions}
from~\citet{optchk-journal-scp-short}, which we adapt to support
higher-order atoms.
A state $S = \exstate{L::G}{\theta}{\emptyset}$,
% where $L$ is a literal,
can be \emph{reduced} to a state $S'$, denoted
$\redA{S}{S'}$, as follows:
\begin{enumerate}
\item If $L$ is a constraint or a higher-order atom, then
  $S' = \exstate{G'}{\theta'}{\emptyset}$, with $G'$ and $\theta'$
  obtained as in the operational semantics without assertions:
  $\red{\state{L::G}{\theta}}{\state{G'}{\theta'}}$.
\item If $L$ is an atom and $\exists(\pred{L}{B})\in\defn{L}$, then
  \[
    S' =
    \begin{cases}
      \exstate{G}{\theta}{\{\aid\}}
      &\hspace*{-2mm}\text{if } \exists C = \acCall{L}{\pre} \in \A
        \quantSep \mathsf{label}(C) = \ell\wedge\notTrivSucc{\pre}\\[1mm]
      \exstate{B :: \mathit{PostC} :: G}{\theta}{\emptyset}
      &\hspace*{-2mm}\text{otherwise}
    \end{cases}
  \]
  where
  $\mathit{PostC} = \checklit{L}{\aid[1]} :: \ldots ::
  \checklit{L}{\aid[n]}$ includes all the checks
  $\checklit{L}{\aid[i]}$ such that $\aid[i] = \mathsf{label}(C_i)$,
  with
  $C_i = \acSucc{L}{\pre_i}{\post_i} \in \A \wedge \trivSucc{\pre_i}$.
\item If $L$ is a check literal $\checklit{L'}{\aid}$, then
  \[
    S' =
    \begin{cases}
      \exstate{G}{\theta}{\{\aid\}}
      &\text{if } \exists C = \acSucc{L}{\_}{\post} \in \A
        \quantSep \mathsf{label}(C) = \ell\wedge\notTrivSucc{\post}\\
      \exstate{G}{\theta}{\emptyset}
      &\text{otherwise}
    \end{cases}
  \]
\end{enumerate}

The set of derivations for a program $\prog$ with assertions $\A$ from
a set of queries $\cqueries$ using the semantics with assertions is
denoted $\derivsA{\prog}{\cqueries}$.
Given a predicate represented by a normalized atom $L$, a store
$\theta$, and a set of queries $\cqueries$, we define the
\emph{success context} $\mathcal{S}_{\A}(L,\theta,\prog,\cqueries)$ of
$L$ and $\theta$ for $\prog$ and $\cqueries$ as
$\{\bar{\exists}_{L}\theta'~|~\exists D \in \derivsA{\prog}{\cqueries}
\quantSep \exists G \quantSep \state{L::G}{\theta} \in D \quantSep
D_{[-1]} = \state{G}{\theta'}\}$.
Intuitively, the \emph{success context} of a predicate $p$ with its
assertions is the set of stores of the success states of $p$ obtained
using the semantics above.
\subsection{Static program analysis by abstract interpretation}
Abstract interpretation \citep{Cousot77-short} is a mathematical
framework for constructing sound, static program analysis tools.
These tools extract properties about a program by interpreting it over
a special \emph{abstract domain} ($\adom$), whose elements are finite
representations of (possibly infinite) sets of actual constraints in
the \emph{concrete domain}~($\cdom$).
Elements of the concrete and abstract domains are related by two
functions: \emph{abstraction} ($\alpha : \cdom \rightarrow \adom$) and
\emph{concretization} ($\gamma : \adom \rightarrow \cdom$).
Provided certain conditions on $\adom$, $\alpha$, and $\gamma$,
abstract interpretation guarantees soundness and termination of the
analysis.
\subsection{Goal-dependent abstract interpretation}
We use, for concreteness, \emph{goal-dependent abstract
  interpretation}, in particular the \plai algorithm~\citep{ai-jlp}.
This technique takes a program $\prog$, an abstract domain $\adom$,
and a set of initial abstract queries $\aqueries$, describing all the
possible initial concrete queries to $\prog$.
An abstract query $\aquery$ is a pair $(L,\lambda)$, where $L$ is an
atom and $\lambda\in\adom$ an abstraction of a set of concrete initial
program states (\emph{e.g.}, constraint stores).
A set of abstract queries $\aqueries$ represents a set of concrete
queries defined as
$\gamma(\aqueries)=\{(L,\theta)~|~(L,\lambda)\in\aqueries\,\wedge\,
\theta\in\gamma(\lambda)\}$.
The algorithm computes a set of triples
$\{\atriple[1],\ldots,\atriple[n]\}$ where $L_i$ is an atom, and
$\lambda_{i}^c$ and $\lambda_{i}^s$ are abstractions approximating the
set of all call and success states for $L_i$, respectively, for all
occurrences of literal $L_i$ in all possible derivations of $\prog$
from $\gamma(\aqueries)$.
Higher-order atoms are supported by reducing them to first-order calls
when the called predicate can be determined by the analysis, or making
conservative assumptions otherwise.
For the rest of the paper, we assume that the abstract
interpretation of a program $\prog$ for the set of initial abstract
queries $\aqueries$, denoted by $\absint$, works with an implicit
abstract domain $\adom$, which safely approximates the concrete values
and operations.
Although not strictly required, $\adom$ has a lattice structure with a
bottom-most element $\bot$, meet $(\sqcap)$, join $(\sqcup)$, and less
than $(\sqsubseteq)$ operators.
As usual, $\bot$ denotes the abstraction s.t.\
$\gamma(\bot)=\emptyset$.
\subsection{Compile-time verification of (first-order) assertions.}
In addition to generating the results mentioned above, the analyzer
also checks any (first-order) assertions in the program by safely
approximating the property formulas of such assertions, and comparing
them against the analysis results ($\absint$) using the abstract
operators.\footnote{We refer the reader
  to~\citet{assrt-theoret-framework-lopstr99-short,verifly-2021-tplp-short}
  for the technical details on this subject.}
The verification result is reported as changes in the status and
transformations of the assertions: \ciaochecked if the properties are
satisfied; \ciaofalse if some property is proved not to hold; or
\ciaocheck if neither of the first two can be determined, in which
case run-time checks will be inserted into the program to ensure
run-time safety.
The verification process yields an assignment of a value
$\ciaochecked,~\ciaofalse,$ or $\ciaocheck$ to each assertion in $\A$,
denoted $\acheck{\A}{\absint}$.
\section{Specifying higher-order programs: predicate properties}
\label{sec:spec-predprops}
In higher-order \clp, variables can be bound to predicate symbols that
are later invoked. This naturally gives rise to the need for
expressing conditions on these predicates that must hold during
program execution.
To this end, \emph{predicate properties} were introduced
in~\citet{asrHO-ppdp2014-shorter}, which we revise and refine
here.\footnote{We propose a more compact syntax here that avoids
  having to use a named variable for the anonymous predicate symbol
  (as in~\citet{asrHO-ppdp2014-shorter}) and takes advantage of
  functional notation (\ciaoinline{:=}).}
A predicate property is defined as a set of \emph{anonymous
  assertions}.
Anonymous assertions generalize traditional assertions by allowing the
predicate symbol in the $\mathit{Pred}$ field to act as a placeholder.

\begin{definition}[Anonymous assertion]
  An \emph{anonymous assertion} $\anonyAssrt$ is an assertion whose
  $\mathit{Pred}$ field is of the form $\anonyHead$, where $\bar{v}$
  are free, distinct variables, and $\_$ is a placeholder for a
  predicate symbol.\footnote{We also use for compactness ``\_'' as
    anonymous functor, a syntactic extension from the \ciao
    \texttt{hiord} package~\citep{ciao-hiord-short}, but double quotes
    \texttt{`'} can also be used to stay within \textsf{ISO-Prolog}
    syntax.}
  Instantiating $\pholder$ with a specific predicate symbol $p$
  produces a traditional assertion for $p$ derived from the anonymous
  assertion $\anonyAssrt$, denoted $\instAnonyAssrt{p}$.
\end{definition}

\begin{example}[Anonymous assertion]
  Let $\anonyAssrt$ be the anonymous assertion:
  ``\ciaoinline{:-$~$pred}~\pholder\ciaoinline{(X,Y)$~$:$~$int(X)$~$=>$~$int(Y).}.''
  Then $\instAnonyAssrt{p}$ is the traditional assertion:
  ``\ciaoinline{:-$~$pred$~$p(X,Y)$~$:$~$int(X)$~$=>$~$int(Y).}''
  obtained by instantiating the anonymous assertion $\anonyAssrt$ with
  the predicate symbol \ciaoinline{p}.
\end{example}

\begin{definition}[Predicate property]
  A \emph{predicate property} $\Pi$ is a set of anonymous assertions
  $\{\seq{\anonyAssrt}\}$.
  Its syntax is:
  ``$\Pi$~\ciaoinline{:=$~$\{$~$:-$~$pred}~$\anonyHead$~\ciaoinline{:}~$\mathit{Pre}_1$~\ciaoinline{=>}~$\mathit{Post}_1$\ciaoinline{.}~~$\ldots$~~\ciaoinline{:-$~$pred}~$\anonyHead$~\ciaoinline{:}~$\mathit{Pre}_n$~\ciaoinline{=>}~$\mathit{Post}_n$\ciaoinline{.$~$\}.}''.
  The function $\ar{\Pi}$ denotes the arity of the predicates for
  which all of the anonymous assertions in $\Pi$ express a property.
  Instantiating $\Pi$ with a specific predicate symbol $p$ produces a
  set of traditional assertions for $p$, denoted
  $\instPredprop{p} =
  \{\instAnonyAssrt[1]{p},\ldots,\instAnonyAssrt[n]{p}\}$.
\end{definition}

We use $\Pi$ to refer to both a set of anonymous assertions and,
interchangeably, the corresponding set of anonymous assertion
conditions; extending \emph{instantiation} accordingly.

\begin{example}[Predicate property]
  The following program defines the predicate property
  \ciaoinline{int_op} (``being a predicate that behaves as an integer
  nondeterministic binary operator''), and a higher-order assertion:
\begin{lstlisting}
int_op := { :- pred _(X,Y,Z) : (int(X), int(Y)) => int(Z). }.
:- pred eval(A,B,$\texttt{Op}$,R) : (int(A), int(B), int_op($\texttt{Op}$)) => int(R).
\end{lstlisting}
  The predicate property literal \ciaoinline{int_op}{\rm\texttt{(Op)}}
  states that {\rm\texttt{Op}} should be a $3$-ary predicate s.t., if
  called with its first two arguments bound to integers, then its
  third argument should be bound to an integer upon success.
  The higher-order assertion for the \ciaoinline{eval/4} predicate
  states that it must be called with its first two arguments bound to
  integers and its third argument bound to a predicate that
  \emph{conforms} to property \ciaoinline{int_op}, and that if any
  such call succeeds, then its fourth argument should be bound to an
  integer.
  \envend
\end{example}
\section{Verifying higher-order programs}
\label{sec:verif}
Once established how to specify higher-order programs using predicate
properties, we now concentrate on how to verify such programs.
We first recall some instrumental definitions
from~\citet{assrt-theoret-framework-lopstr99-short} for reasoning
about abstractions of property formulas.
For the rest of the discussion, let $\prog$ be a program and $F$ a
property formula defined in $\prog$.
\begin{definition}[Trivial success set]
  \label{def:trivial-suc-set}
  We define the \emph{trivial success set} of $F$ as
  $\trivset{F} = \{ \bar{\exists}_F\theta \, |\, \trivSucc{F} \}$.
\end{definition}
\begin{example}[Trivial success set]
  Let $F=$ \ciaoinline{list(L)}, both $\theta_1=\{$\ciaoinline{L
    =$~$[1,2]}$\}$ and $\theta_2=\{$\ciaoinline{L =$~$[1,X]}$\}$ are
  in $\trivset{F}$, but $\theta_3=\{$\ciaoinline{L =$~$[1|_]}$\}$ is
  not, since a call to \ciaoinline{(L =$~$[1|_], list(L))} would
  further instantiate the second argument of \ciaoinline{[1|_]}.
  The trivial success set $\trivset{F}$ of $F$ captures the notion
  that the \ciaoinline{list(L)} property formula requires \texttt{L}
  to be \emph{instantiated} to (the structure of) a list.
\end{example}
\begin{definition}[Abstract trivial success subset]
  \label{def:abstract-trivial-sub-set}
  An abstraction is an \emph{abstract trivial success subset} of $F$,
  denoted $\abstrivsubset{F}{\prog}$, iff
  $\gamma(\abstrivsubset{F}{\prog}) \subseteq \trivset{F}$.
\end{definition}
\begin{definition}[Abstract trivial success superset]
  \label{def:abstract-trivial-sup-set}
  An abstraction is an \emph{abstract trivial success superset} of
  $F$, denoted $\abstrivsupset{F}{}$, iff
  $\gamma(\abstrivsupset{F}{\prog}) \supseteq \trivset{F}$.
\end{definition}

Intuitively, $\abstrivsubset{F}{\prog}$ and $\abstrivsupset{F}{\prog}$
are, respectively, a safe under- and over-approximation of the trivial
success set $\trivset{F}$ of the property formula $F$, and they can
always be computed at compile-time by choosing the closest element in
the abstract domain.
\subsection{Conformance to a predicate property}
\label{sec:predprop_conf}
When we provide a partial specification for a higher-order argument
$X$ of a higher-order predicate using a predicate property $\Pi$, we
are describing requirements that the predicates that $X$ may be bound
to must meet.
We refer to a predicate $p$ behaving correctly w.r.t.~$\Pi$ as: $p$
\emph{conforming} to $\Pi$.
We will now formalize this notion, with the goal of being able to
safely approximate the set of predicates that $X$ can be bound to
without violating $\Pi$.
\begin{definition}[Covered predicate]
  \label{def:covered}
  Given $C = \acCall{\anonyHead}{\anonyPre} \in \Pi$ and
  $\anonyACond = \acCall{\p}{\mathit{Pre}} \in \A$, we say that $p$
  can be covered with $\Pi$ iff
  $\trivset{\anonyPre} \subseteq \trivset{\pre}$.
\end{definition}

Intuitively, $p$ can be covered with $\Pi$ if the set of admissible
calls to $p$ is a superset of the set of admissible calls described by
$\Pi$.

\begin{definition}[Redundance]
  \label{def:redundance}
  Under the same conditions as in~\cref{def:covered}, given that $p$
  \emph{can be covered with} $\Pi$, we define the set of assertion
  conditions $\A'$ as follows:
  \[
    \A' = \{\acCall{\p}{\mathit{Pre} \wedge \anonyPre}\} \cup (\A
    \setminus \{C\}) \cup (\Pi \setminus \{\anonyACond\})\hspace{-0.1em}|_{p}
  \]
  Given a sequence of literals $G$, let $\mathcal{U}(G)$ denote the
  result of removing all check literals from $G$. We extend
  $\mathcal{U}$ to derivations so that $\mathcal{U}(D)$ denotes the
  derivation resulting from transforming any (extended) state
  $\exstate{G}{\theta}{\falseAc}$ in $D$ into the state
  $\state{G'}{\theta}$, where $G'= \mathcal{U}(G)$.
  Let $\cquery_p$ be a query to $p$.
  We say that $\Pi$ is \emph{redundant} for $p$ under $\cquery_p$ iff
  \[
    \forall D' \in \derivsA[\A']{\prog}{\cquery_p} \quantSep D'_{[-1]} =
    \exstate{G'}{\theta}{\{\aid'\}},
  \]
  and
  \[
    \forall D \in \derivsA{\prog}{\cquery_p} \quantSep
    \mathcal{U}(D) = \mathcal{U}(D'),
  \]
  it holds that
  $\redstarA{D_{[-1]}}{\exstate{G}{\theta}{\{\aid\}}}$ through a
  derivation that reduces only check literals (if any at
  all),\footnote{Note that this implies
    $\mathcal{U}(G) = \mathcal{U}(G')$.}
  where $\aid$ (resp., $\aid'$) is the label for a $\mathsf{calls}$ or
  $\mathsf{success}$ assertion condition in $\A(\p)$
  (resp., $\A'(\p)$).
\end{definition}

Intuitively, a predicate property $\Pi$ is redundant for a predicate
$p$ under a query $\cquery_p$ to $p$ iff augmenting the original set
of assertion conditions ($\A$) with that of $\Pi$ ($\A'$) does not
introduce new run-time check errors in any derivation starting from
$\cquery_p$.\footnote{Note that, for the purposes of
  determining conformance, the assertions for the predicates in the
  program can be provided by the user, inferred by analysis, or a
  combination of both.}
\ \\[1mm]
\begin{definition}[Conformance]
  \label{def:conformance}
  Let $\cqueries_p$ be the set of all possible queries to $p$.
  A predicate $p$ \emph{conforms} to $\Pi$, denoted $p \conf \Pi$, iff
  $\forall \cquery_p \in \cqueries_p \quantSep \Pi$ is redundant for
  $p$ under $\cquery_p$.
  Conversely, $p$ \emph{does not conform} to $\Pi$, denoted
  $p \nconf \Pi$, iff
  $\exists \cquery_p \in \cqueries_p \quantSep \Pi$ is not redundant
  for $p$ under $\cquery_p$.
\end{definition}

To prove that a predicate conforms to a predicate property, all
possible derivations from all possible queries to that predicate have
to be considered, which is often not feasible in practice.
To this end, we introduce the notion of \emph{abstract conformance} as
a compile-time conformance criterion.
Abstract conformance safely approximates the notion of conformance by
comparing the assertion conditions of a predicate and those of a
predicate property under the order relation of an abstract domain.
We denote by $(\aconf)$ the notion of \emph{strong} abstract
conformance, and by $(\aconfover)$ that of \emph{weak} abstract
conformance.
That is, an under- and over-approximation of abstract conformance,
respectively.
Intuitively, strong abstract conformance captures only predicates
known to conform, while weak abstract conformance also includes those
for which conformance is unknown.
Thus, the negation of weak abstract conformance captures the
predicates that are known not to conform.

\begin{definition}[Abstract conformance on ``calls'']
  \label{def:conf-call}
  Let $\pre$ be the pre-condition of the $\mathsf{calls}$ assertion
  condition for $p$ in $\A$, and $\anonyACond$ be an anonymous
  $\mathsf{calls}$ assertion condition
  $\acCall{\anonyHead}{\anonyPre}$.
  Then:
  \begin{align*}
    p \aconf  \anonyACond
    &\Leftrightarrow
      (\abstrivsupset{\pre}{\prog}
       \sqsubseteq
       \abstrivsubset{\anonyPre}{\prog})
      \wedge
      (\abstrivsubset{\pre}{\prog}
       \sqsupseteq
       \abstrivsupset{\anonyPre}{\prog})\\
    p \naconf \anonyACond
    &\Leftrightarrow
      \abstrivsupset{\pre}{\prog}
      \sqcap
      \abstrivsupset{\anonyPre}{\prog}
      = \bot
  \end{align*}
\end{definition}
\begin{definition}[Abstract conformance on ``success'']
  \label{def:conf-succ}
  Let $\A$ be the set of assertion conditions for $p$, 
  and $\anonyACond$ be an anonymous $\mathsf{success}$ assertion
  condition $\acSucc{\anonyHead}{\anonyPre}{\anonyPost}$.
  Then:
  \begin{align*}
    p \aconf \anonyACond \Leftrightarrow
    &\;\exists S \subset \A,
      (\abstrivsubset{\pre}{\prog}_{\sqcup}
       \sqsupseteq
       \abstrivsupset{\anonyPre}{\prog})
      \wedge
      (\abstrivsupset{\post}{\prog}_{\sqcup}
       \sqsubseteq
       \abstrivsubset{\anonyPost}{\prog}),\\[-.5mm]
    &\;\text{ where }
      \begin{cases}
        \abstrivsubset{\pre}{}_{\sqcup}
        & \hspace*{-.6em}=\sqcup\{\abstrivsubset{\pre}{}~~\hspace*{.3mm}|~\acSucc{\p}{\pre}{\_} \;\;\hspace{.3mm}\in S\}\\[-.5mm]
        \abstrivsupset{\post}{}_{\sqcup}
        & \hspace*{-.6em}=\sqcup\{\abstrivsupset{\post}{}~|~\acSucc{\p}{\_}{\post} \, \in S\}
      \end{cases}\\
    p \naconf \anonyACond \Leftrightarrow
    &\;\exists\,\acSucc{\p}{\pre}{\post}\in \A,
    (\abstrivsupset{\mathit{Pre}}{\prog}
     \sqsubseteq
     \abstrivsubset{\anonyPre}{\prog})
     ~\wedge\\[-1mm]
    &\wedge
      (\abstrivsupset{\post}{\prog}
       \sqcap
       \abstrivsupset{\anonyPost}{\prog}
      = \bot)
\wedge \exists \theta \in \trivset{\pre} \quantSep
    \mathcal{S}_{\A}(\p,\theta,\prog,\gamma(\aqueries_{p})) \neq \emptyset
  \end{align*}
  where $\aqueries_p$ is the set of abstract queries
  s.t.~$\gamma(\aqueries_p)$ is a \emph{superset} of the set of all
  valid queries to $p$ described by the $\mathsf{calls}$ assertion
  condition of $p$ in $\A$.
\end{definition}
\begin{definition}[Abstract conformance]
  \label{def:abs-conf}
  We define abstract conformance to a predicate property as follows:
  \begin{align*}
    p \aconf \Pi
    &\Leftrightarrow \forall\,\anonyACond \in \Pi_C \quantSep p \aconf \anonyACond\\
    p \naconf \Pi
    &\Leftrightarrow \exists\,\anonyACond \in \Pi_C \quantSep p \naconf \anonyACond
  \end{align*}
\end{definition}

Note that abstract conformance is computed by first computing the
abstract trivial success subsets or supersets of the involved property
formulas, and then applying the operators of the abstract domain.
For abstract domains which may lose precision with their ($\sqcup$)
abstract operator, more advanced techniques for leveraging multiple
abstractions become necessary, \emph{e.g.},
\emph{covering}~\citep{non-failure-iclp97}.
We now relate the notions of conformance and abstract conformance.
\begin{theorem}
  \label{th:absconf}
  Let $p$ be a predicate, $\Pi$ be a predicate property:
  $p \aconf \Pi \Rightarrow p \conf \Pi$, and
  $p \naconf \Pi \Rightarrow p \nconf \Pi$.
\end{theorem}
\begin{proof}[Proof.]
The proofs proceed by contradiction and direct proof,
respectively, using
Defs.~\labelcref{def:covered,def:redundance,def:conformance,def:abstract-trivial-sub-set,def:abstract-trivial-sup-set,def:conf-call,def:conf-succ}
and some basic set manipulation.
Detailed proofs can be found in
Appendix~\hyperref[app:proof:conf]{A}.\hfill
\end{proof}
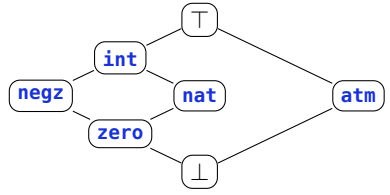
\begin{figure}
  \begin{flushleft}
  \begin{subfigure}{.652\textwidth}
\begin{lstlisting}[language=CiaoLong,basicstyle=\footnotesize\ttfamily]
p_nat_nat := { :- pred _(X,Y) : nat(X) => nat(Y). }.
\end{lstlisting}
    \vspace{-2mm}
    \caption{Predicate property.}
    \label{fig:case-analysis-predprop}
    \vspace{-1mm}
  \end{subfigure}
  \end{flushleft}
  \begin{minipage}{.56\textwidth}
    \begin{subfigure}{\textwidth}
      \centering
\begin{lstlisting}[firstnumber=2,language=CiaoLong,basicstyle=\footnotesize\ttfamily]
:- pred  n2n(X,Y) :  nat(X) =>  nat(Y). % $\color{ciaofalse}A_1$
:- pred  a2n(X,Y) :  atm(X) =>  nat(Y). % $\color{ciaofalse}A_2$
:- pred  i2z(X,Y) :  int(X) => zero(Y). % $\color{ciaofalse}A_3$
:- pred  z2i(X,Y) : zero(X) =>  int(Y). % $\color{ciaofalse}A_4$
:- pred nz2n(X,Y) : negz(X) =>  nat(Y). % $\color{ciaofalse}A_5$
\end{lstlisting}
      \vspace{-2mm}
      \caption{Assertions.}
      \label{fig:case-analysis-source-calls}
    \end{subfigure}
  \end{minipage}
  \hfill
  \begin{minipage}{.4\textwidth}
    \begin{subfigure}{\textwidth}
      \centering
      \vspace{-6mm}
      \begin{tikzpicture}[xscale=.7]
        \node[draw, rounded corners] (top)  at (   0,   3) {$\top$};
        
        \node[draw, rounded corners] (int)  at (-1.5, 2.5) {\ciaoinline{int}};
        
        \node[draw, rounded corners] (negz) at (  -3,   2) {\ciaoinline{negz}};
        \node[draw, rounded corners] (nat)  at (   0,   2) {\ciaoinline{nat}};
        \node[draw, rounded corners] (atm)  at (   3,   2) {\ciaoinline{atm}};
        
        \node[draw, rounded corners] (zero) at (-1.5, 1.5) {\ciaoinline{zero}};
        
        \node[draw, rounded corners] (bot)  at (   0,   1) {$\bot$};
        
        \draw
        (top)  -- (int)
        (top)  -- (atm)
        
        (int)  -- (negz)
        (int)  -- (nat)
        (atm)  -- (bot)
        
        (negz) -- (zero)
        (nat)  -- (zero)
        
        (zero) -- (bot);
      \end{tikzpicture}
      \caption{\centering Abstract domain lattice.}
      \label{fig:case-analysis-lattice}
    \end{subfigure}
  \end{minipage}
  \caption{\centering Example case analysis on a predicate property and assertions.}
\end{figure}
\begin{example}[Abstract conformance]
  Consider determining conformance to the predicate property
  in~\cref{fig:case-analysis-predprop} -- which, for simplicity, we
  will interchangeably refer to as $\Pi$ for the rest of the
  example -- given the assertions $\A=\{\seq[5]{A}\}$
  in~\cref{fig:case-analysis-source-calls}.
  (Notice that the property formulas of both $\Pi$ and $\A$ include
  elements of the abstract domain represented by the lattice
  in~\cref{fig:case-analysis-lattice}).
  Their corresponding sets of assertion conditions are:
  \begin{align*}
    \Pi &= \{
            \acCall{\texttt{\_(X,Y)}}{\text{~\ciaoinline{nat(X)}}},~
            \acSucc{\texttt{\_(X,Y)}}{\text{~\ciaoinline{nat(X)}}}{\text{~\ciaoinline{nat(Y)}}}\}\\
    \A &= \{
          \acCall{\mathit{Pred}_i}{\mathit{Pre}_i},~
          \acSucc{\mathit{Pred}_i}{\mathit{Pre}_i}{\mathit{Post}_i}\; |\, A_i \in \A\}
  \end{align*}
  We aim to determine which predicates partially specified by $\A$
  abstractly conform to the predicate property $\Pi$.
  For each predicate $\mathit{Pred}_i$ and its associated
  $\mathsf{calls}$ and $\mathsf{success}$ assertion conditions, we:
  (1) determine abstract conformance to the anonymous $\mathsf{calls}$
  condition of $\Pi$; (2) determine abstract conformance to the
  anonymous $\mathsf{success}$ condition of $\Pi$; and (3) determine
  abstract conformance to $\Pi$:

  \cref{tb:calls,tb:success} summarize the abstract conformance
  analysis between the $\mathsf{calls}$ and $\mathsf{success}$
  assertion conditions of each predicate and those of $\Pi$,
  respectively.
  Specifically, for $\mathsf{calls}$ (in~\cref{tb:calls}), we compare
  the pre-conditions and apply~\cref{def:conf-call}; for
  $\mathsf{success}$ (in~\cref{tb:success}), we compare both pre- and
  post-conditions and apply~\cref{def:conf-succ}.
  \begin{table}
    \centering
    \caption{{\itshape Abs.~conf.~on ``\emph{calls}'' example with} $\anonyPre=$ \ciaoinline{nat(X)}.}
    \label{tb:calls}
    {\begin{minipage}{.92\textwidth}\centering\tablefont\small\begin{tabular}{@{\extracolsep{\fill}}lccc}\midrule\midrule
      $\bm{\mathit{Pred}_i}$
      & $\bm{\mathit{Pre}_i}$
      & \textbf{Relation w.} $\anonyPreBold$
      & \textbf{Abs.\ Conf.}\\\midrule
      \ciaoinline{n2n(X,Y)}
      & \ciaoinline{nat(X)}
      & \ciaoinline{nat(X)} $=$ \ciaoinline{nat(X)}
      & yes\\
      \ciaoinline{a2n(X,Y)}
      & \ciaoinline{atm(X)}
      & \ciaoinline{atm(X)} $\sqcap$ \ciaoinline{nat(X)} $=\bot$
      & no\\
      \ciaoinline{i2z(X,Y)}
      & \ciaoinline{int(X)}
      & \ciaoinline{int(X)} $\sqsupseteq$ \ciaoinline{nat(X)}
      & maybe\\
      \ciaoinline{z2i(X,Y)}
      & \ciaoinline{zero(X)}
      & \ciaoinline{zero(X)} $\sqsubseteq$ \ciaoinline{nat(X)}
      & maybe\\
      \ciaoinline{nz2n(X,Y)}
      & \ciaoinline{negz(X)}
      & \ciaoinline{negz(X)} $\sqcap$ \ciaoinline{nat(X)} $\not=\bot$
      & maybe\\
      & & $\wedge$ \ciaoinline{negz(X)} $\not\sqsubseteq$ \ciaoinline{nat(X)}\\
      & & $\wedge$ \ciaoinline{negz(X)} $\not\sqsupseteq$ \ciaoinline{nat(X)}\\\midrule\midrule
    \end{tabular}\end{minipage}}
  \end{table}
  \begin{table}
    \centering
    \caption{Abs.~conf.~on ``\emph{success}'' example with $\anonyPre=$ \ciaoinline{nat(X)} and $\anonyPost=$ \ciaoinline{nat(Y)}.}
    \label{tb:success}
    {\begin{minipage}{.92\textwidth}\centering\tablefont\small\begin{tabular}{@{\extracolsep{\fill}}lccccc}\midrule\midrule
      $\bm{\mathit{Pred}_i}$
      & $\bm{\mathit{Pre}_i}$
      & \textbf{Relation w.} $\anonyPreBold$
      & $\bm{\mathit{Post}_i}$
      & \textbf{Relation w.} $\anonyPostBold$
      & \textbf{Abs.\ Conf.}\\\midrule
      \ciaoinline{n2n(X,Y)}
      & \ciaoinline{nat(X)}
      & \ciaoinline{nat(X)} $=$ \ciaoinline{nat(X)}
      & \ciaoinline{nat(Y)}
      & \ciaoinline{nat(Y)} $=$ \ciaoinline{nat(Y)}
      & yes\\
      \ciaoinline{a2n(X,Y)}
      & \ciaoinline{atm(X)}
      & \ciaoinline{atm(X)} $\sqcap$ \ciaoinline{nat(X)} $=\bot$
      & \ciaoinline{nat(Y)}
      & \ciaoinline{nat(Y)} $=$ \ciaoinline{nat(Y)}
      & maybe\\
      \ciaoinline{i2z(X,Y)}
      & \ciaoinline{int(X)}
      & \ciaoinline{int(X)} $\sqsupseteq$ \ciaoinline{nat(X)}
      & \ciaoinline{zero(Y)}
      & \ciaoinline{zero(Y)} $\sqsubseteq$ \ciaoinline{nat(Y)}
      & yes\\
      \ciaoinline{z2i(X,Y)}
      & \ciaoinline{zero(X)}
      & \ciaoinline{zero(X)} $\sqsubseteq$ \ciaoinline{nat(X)}
      & \ciaoinline{int(Y)}
      & \ciaoinline{int(Y)} $\sqsupseteq$ \ciaoinline{nat(Y)}
      & maybe\\
      \ciaoinline{nz2n(X,Y)}
      & \ciaoinline{negz(X)}
      & \ciaoinline{negz(X)} $\sqcap$ \ciaoinline{nat(X)} $\not=\bot$
      & \ciaoinline{nat(Y)}
      & \ciaoinline{nat(Y)} $=$ \ciaoinline{nat(Y)}
      & maybe\\
      & & $\wedge$ \ciaoinline{negz(X)} $\not\sqsubseteq$ \ciaoinline{nat(X)} & & &\\
      & & $\wedge$ \ciaoinline{negz(X)} $\not\sqsupseteq$ \ciaoinline{nat(X)} & & &\\\midrule\midrule
    \end{tabular}\end{minipage}}
  \end{table}
  As a summary, the only predicate that definitely conforms to
  \ciaoinline{p_nat_nat} is \ciaoinline{n2n/2}, since both of its
  assertion conditions conform to $\Pi$.
\end{example}
\subsection{Wrappers}
\label{sec:wrappers}
Consider a predicate $p$ and a predicate property $\Pi$ s.t.~$p$ can
be covered by $\Pi$.
From~\cref{def:covered} we know that given their respective
pre-conditions $\mathit{Pre}$ and $\anonyPre$,
$\trivset{\pre} \supseteq \trivset{\anonyPre}$.
Thus, according to~\cref{def:conf-call}, $p$ \emph{may} abstractly
conform to $\Pi$ ($p \aconfover \Pi$), since $\pre$ may describe
\emph{more} admissible call states for $p$ than $\anonyPre$, which can
lead to omitting some run-time check errors that would be raised by
$\anonyPre$.

\begin{example}[Weak abstract conformance]
  Consider a query \ciaoinline{?-$~$foo(even)} to the following
  program.
\begin{lstlisting}
p_nat := { :- pred _(N) : nat(N). }. % 1(*\color{ciaofalse}-*)ary predicates for naturals(*\color{ciaofalse}.*)

:- pred even(N) : int(N).            :- pred foo(P) : p_nat(P).
even(N) :-                           foo(P) :- P(10).  % (*\color{ciaofalse}$(1)$\label{line1}*)
    integer(N), 0 is N mod 2.        foo(P) :- P(-10). % (*\color{ciaofalse}$(2)$\label{line2}*)
\end{lstlisting}
  Take a derivation of such query that starts by reducing to the body
  of the first clause $(1)$: no $\mathsf{calls}$ assertion condition
  violation is expected since all predicates that conform to
  \ciaoinline{p_nat} \emph{must} accept all natural numbers on calls.
  Now, take a derivation that reduces to the body of the second clause
  $(2)$: a $\mathsf{calls}$ assertion condition violation is expected
  since all predicates that conform to \ciaoinline{p_nat} \emph{must}
  raise an error for any input different from a natural number.
  However, in this particular case, no error is raised, since the
  predicate \texttt{even/1} accepts any integer on calls.
  Moreover, if we had:
\begin{lstlisting}[firstnumber=6]
p_neg := { :- pred _(N) : neg(N). }. % 1(*\color{ciaofalse}-*)ary predicates for negatives(*\color{ciaofalse}.*)

:- pred bar(P) : p_neg(P).
bar(P) :- P(-4).
\end{lstlisting}
  then a clause like
  \ciaoinline{foo(P)}{\tt~}\ciaoinline{:-}{\tt~}\ciaoinline{bar(P).}
  would be problematic.
  For this clause, looking at the assertion of \ciaoinline{foo(P)}, the
  predicate in \ciaoinline{P} is required to conform to
  \ciaoinline{p_nat}, which would be an error when calling
  \ciaoinline{bar(P)} since \ciaoinline{p_neg} is \emph{disjoint} from
  \ciaoinline{p_nat}.
  However, if we consider the particular case in which \ciaoinline{P
    =}{\tt~}\ciaoinline{even}, it may not.
  So, according to~\cref{def:conf-call}, we could only conclude that
  \ciaoinline{even/1} $\aconfover$ \ciaoinline{p_nat}, that is,
  \ciaoinline{even/1} \emph{may} abstractly conform.
\end{example}

In the example above, we motivate the need for such a restrictive
condition for abstract conformance on \emph{calls}
(see~\cref{def:conf-call}).
However, we may want to use predicates whose set of admissible calls
is greater than that of a predicate property, but without unexpected
behavior.
To this end, we propose a technique to restrict the set of admissible
calls of a predicate $p$ described by $\pre$ to match that of
$\anonyPre$ in a program analysis friendly manner.
This restriction is implemented using \emph{wrappers}.
A wrapper for $p$ with $\Pi$ is simply a new predicate
$\pred{w(\bar{v})}{p(\bar{v})}$ with an assertion
``\ciaoinline{:-$~$pred} $w(\bar{v})$ \ciaoinline{:}
$\anonyPre$\ciaoinline{.}''  (note that fields of \emph{pred}
assertions, in this case the post-condition, can be omitted,
equivalently to \emph{true}).
A wrapper for $p$ with $\Pi$ also makes explicit the intention of
creating a $\Pi$-tailored version of $p$.
Additionally, wrappers can also be used to alleviate the process of
determining abstract conformance on \emph{calls} (particularly useful
in the implementation), since the wrapper would syntactically (and
thus, semantically) match the pre-condition of the predicate property.

\begin{example}[Wrapper]
  As a follow-up of the previous example, consider \emph{wrapping}
  \ciaoinline{even/1} with \ciaoinline{p_nat}:
\begin{lstlisting}[firstnumber=10]
:- pred even_nat(N) : nat(N).
even_nat(N) :- even(N).
\end{lstlisting}
  Intuitively, \ciaoinline{even_nat/1} conforms to \ciaoinline{p_nat},
  and the analyzer can now infer that the clause
  \ciaoinline{foo(P)}{\tt~}\ciaoinline{:-}{\tt~}\ciaoinline{bar(P).}
  should raise an error since \ciaoinline{even_nat/1} only accepts
  naturals.
\end{example}
\subsubsection{Rationale for explicit wrappers}
The design of {\itshape\hiord}~follows the philosophy behind the \ciao
system~\citep{ciao-design-tplp-shorter}, which extends \textsf{Prolog}
with static and dynamic assertion checking (among other modular
extensions) without altering its untyped nature.
We also considered some alternative solutions to the problem at
hand, such as \emph{tainting} each predicate passed as an argument
annotated with a \emph{predicate property}, and restricting its future
use in all internal (recursive) calls.
However, this approach would have required modifying the standard
\textsf{Prolog} semantics regarding higher-order calls.
The use of \emph{wrappers} allows us to simulate this behavior
without altering the underlying semantics.
\subsection{First-order representation of predicate properties}
\label{sec:ground-predprops}
As mentioned in~\cref{sec:prel}, the abstract interpretation-based
static analyzer can \emph{infer} properties about higher-order
programs, and also \emph{verify} first-order assertions.
However, here we obviously need to deal with predicate properties in
assertions.
Usually, for a new type of property, a new abstract domain is needed.
As an alternative approach, we herein propose representing predicate
properties as first-order properties of a kind which can be natively
supported by the analyzer, thus allowing us to leverage existing and
mature abstract domains.
More concretely, we propose representing predicate properties as
\emph{regular types}, a special kind of properties (and thus defined
as predicates) that are used to describe the \emph{shape} of a term.
Intuitively, such types will capture sets of predicate names.
For example, given the predicate property \ciaoinline{pp}, we can
represent that the predicates \texttt{p}, and \texttt{q} strongly, and
\texttt{r} weakly conforms to \ciaoinline{pp} as the following regular
types: \ciaoinline{pp}$^{-}$\ciaoinline{/1} =
$\{$\ciaoinline{pp}$^{-}$\ciaoinline{(p)},
\ciaoinline{pp}$^{-}$\ciaoinline{(q)}$\}$ and
\ciaoinline{pp}$^{+}$\ciaoinline{/1} =
$\{$\ciaoinline{pp}$^{+}$\ciaoinline{(p)},
\ciaoinline{pp}$^{+}$\ciaoinline{(q)},
\ciaoinline{pp}$^{+}$\ciaoinline{(r)}$\}$.\footnote{Or, using \ciao's
  functional notation:
  ``\ciaoinline{:-$~$regtype$~$pp}$^{+}$\ciaoinline{/1.$~$$~$pp}$^{+}$\ciaoinline{$~$:=$~$p$~$|$~$q$~$|$~$r.}.''}
Formally, given a predicate property
$\Pi$, we define two associated regular types:
$\pi^{-}$\texttt{/1} and
$\pi^{+}$\texttt{/1}, that capture the set of predicates that
\emph{strongly} and \emph{weakly} abstractly conform to
$\Pi$ as follows: $\pi^{-}\texttt{/1} = \{\pi^{-}(p)~|~p \in \prog
\wedge p \aconf \Pi\}$, and $\pi^{+}\texttt{/1} = \{\pi^{+}(p)~|~p \in
\prog \wedge p \aconfover \Pi\}$.
By definition,
$\pi^{-}$\texttt{/1} is a \emph{subtype} of $\pi^{+}$\texttt{/1}.
These regular types reduce the compile-time checking of higher-order
assertions to that of first-order assertions.
Regular types can be abstracted and inferred by several abstract
domains; for concreteness we use
\textsf{eterms}~\citep{eterms-sas02-short}.
\subsection{\hiord~algorithm}
\label{sec:hiord-algo}
\begin{figure}[t]
  \vspace{-2mm}
\begin{algorithm}[H]
  \caption{[{\itshape\hiord}]: Verify a higher-order program with higher-order assertions}
  \label{alg:hiord}
  \textbf{Input:} Program: $\prog$, assertions: $\A$, abstract queries: $\aqueries$\\
  \textbf{Output:} Verified status (\ciaochecked/\ciaofalse/\ciaocheck) for the assertions $\A$ of $\prog$: $\mathcal{V}$
  \vspace*{-1mm}
  \begin{algorithmic}[1]
    \State $R \gets \emptyset$
    \Repeat \hfill $\triangleright$ \emph{start fixpoint computation}\label{alg:line2}
      \State $R' \gets R$ \hfill $\triangleright$ \emph{save state to check fixpoint convergence}
      \ForAll{predicate property $\Pi \in \prog$}\label{alg:line4}
        \State $R \gets R \cup \{\pi^{-}(p)~|~p \in \prog \wedge p\aconf\Pi\}\cup\{\pi^{+}(p)~|~p \in \prog \wedge p\aconfover\Pi\}$\
        \hfill $\triangleright$ \emph{regtypes}\label{alg:line5}
      \EndFor\label{alg:line6}
    \Until{$R = R'$} \hfill $\triangleright$ \emph{fixpoint reached}\label{alg:line7}
    \State $\mathcal{V} \gets \acheck{\A}{\absint[\prog \cup R]}$
      \hfill $\triangleright$ \emph{first-order assertion checking process}\label{alg:line8}
  \end{algorithmic}
\end{algorithm}
\vspace*{-2mm}
\end{figure}
We now present {\itshape\hiord}, the core algorithm for the
compile-time verification of a higher-order program
$\prog$ with higher-order assertions $\A$ (\cref{alg:hiord}).
First, it initializes a set of rules
$R$, and it computes the regular type representations of each
predicate property $\Pi$ in $\prog$, that is, the $\pi^{-}$ and
$\pi^{+}$ predicates respectively (lines~\ref{alg:line4}
to~\ref{alg:line6}).
This computation is performed by directly applying
\cref{def:conf-call,def:conf-succ,def:abs-conf} using the operators of
the abstract domain, and (implicitly) extending the program
$\prog$ with $R$.
Since predicate properties can include predicate property literals
from other predicate properties -- that is, \emph{dependencies}
among predicate properties -- lines~\ref{alg:line4} to~\ref{alg:line6}
are repeated until a \emph{fixpoint} is reached (lines~\ref{alg:line2}
and~\ref{alg:line7}).
Next, it computes the abstract interpretation of
$\prog$, augmented with the regular type representations of every
predicate property, for the set of abstract queries
$\aqueries$ (line~\ref{alg:line8}).
Finally, it performs the compile-time verification of the set of (now
first-order) assertions
$\A$ w.r.t.~the static analysis results, where predicate properties
are now treated as standard regular types (line~\ref{alg:line8}).
As the result of the algorithm, we obtain the verified status of each
assertion of
$\A$, where each assertion can be discharged (\ciaochecked), disproved
(\ciaofalse) and an error flagged, or left in \ciaocheck status, and
subject to run-time checks, as in~\cite{asrHO-ppdp2014-shorter}.
We argue that, despite the inherent complexity of the verification
problem in hand, the proposed concepts make the compile-time checking
algorithm clear and concise; and, more importantly, easily
implementable using a first-order assertion checker.
\section{Implementation and experiments}
\label{sec:exp}
\lstset{language=CiaoLong,basicstyle=\footnotesize\ttfamily}
To demonstrate the potential of our approach, we have implemented a
prototype of the {\itshape\hiord} technique as part of the \ciao\
system.
It implements \cref{alg:hiord} and uses \ciaopp, the \ciao program
preprocessor, with the \textsf{eterms} abstract domain.
We ran experiments on a set of small but representative higher-order
programs that were not possible to verify until this point.
We illustrate below our experiments with a selection of these
programs.
\subsection{A synthetic benchmark}
We started by defining a test case comprising a predicate property
using an anonymous \emph{pred} assertion
and 25 predicates, each with a \emph{pred} assertion, designed to
exhaustively cover all possible orderings between the pre- and
post-conditions of the predicate property and of each predicate.
We then ran {\itshape\hiord}, obtaining the correct results that 2
predicates \emph{definitely did conform} and 7 predicates
\emph{definitely did not conform}, with 16 predicates left where no
definite conclusion could be reached.
\subsection{Higher-order list utilities}
We defined various partially specified higher-order utility predicates
specialized for working with lists of a particular type
\ciaoinline{t}, for example,
\ciaoinline{t(X)}$\,$\ciaoinline{:-}$\; $\ciaoinline{num(X)}.
For example, consider the \ciaoinline{t_cmp} predicate property
defined below:
\begin{lstlisting}
t_cmp := { :- pred _(X,Y) : (t(X), t(Y)). }.
\end{lstlisting}
which describes \emph{comparator} predicates of elements of type
\ciaoinline{t}, that we then use in the higher-order assertion for a
comparator-parameterizable \emph{quicksort} implementation:
\begin{lstlisting}[firstnumber=2]
:- pred qsort(Xs,P,Ys) : (list(t,Xs), t_cmp(P)) => list(t,Ys).
\end{lstlisting}
and where the \texttt{partition/4} predicate includes a call
\texttt{P(X,Y)}.
The analysis is able to propagate the \ciaoinline{t_cmp} property on
\texttt{P} to that point,
and if in \texttt{P(X,Y)}, \texttt{X} is inferred to be bound to,
for example, \texttt{a}, an error is statically captured by {\itshape\hiord}.
Consider the following comparators:

\noindent\begin{minipage}{.48\textwidth}
\begin{lstlisting}[firstnumber=3]
:- pred lex(X,Y) : (term(X), term(Y)).
lex(X,Y) :- X @< Y.
\end{lstlisting}
\end{minipage}
\hfill
\begin{minipage}{.48\textwidth}
\begin{lstlisting}[firstnumber=5]
:- pred lex_t(X,Y) : (t(X), t(Y)).
lex_t(X,Y) :- lex(X,Y).
\end{lstlisting}
\end{minipage}

\noindent For a query \ciaoinline{?-$~$qsort(Xs,lex,Ys)},
{\itshape\hiord} reports a warning on \emph{calls} since
\ciaoinline{lex/2} $\aconfover$ \ciaoinline{t_cmp} (it weakly
abstractly conforms).
Intuitively, \ciaoinline{lex/2} is not definitely conformant since it
will not raise a run-time check error when called with a term that is
not of type \ciaoinline{t}, introducing unexpected behavior.
For a query \ciaoinline{?-$~$qsort(Xs,lex_t,Ys)} {\itshape\hiord}
proves that it behaves correctly w.r.t.~its higher-order assertion,
since \ciaoinline{lex_t/2} $\aconf$ \ciaoinline{t_cmp}.

Additionally, consider a predicate property which represents a
parameterizable sorter of lists of elements of type \ciaoinline{t},
defined ``in terms of'' the \ciaoinline{t_cmp} predicate property:
\begin{lstlisting}[firstnumber=7]
t_sort := { :- pred _(Xs,C,Ys) : (list(t,Xs), t_cmp(C)) => list(t,Ys). }.
\end{lstlisting}
For determining that \texttt{qsort/3} $\aconf$ \ciaoinline{t_sort},
{\itshape\hiord} would need to perform an additional iteration of the
fixpoint computation after the one above, that is, after computing the
predicates that weakly or strongly abstractly conform to the
\ciaoinline{t_cmp} predicate property.
\subsection{\textsf{HTTP} server}
Consider the following schematic \textsf{HTTP} server, parameterized
by a predicate that must be able to handle four \textsf{REST}
operations.
We use regular types for representing requests and responses, and a
predicate property for representing handlers:
\begin{lstlisting}
handler := { :- pred _(Rq,Rs) : req(Rq) => res(Rs). }.

:- regtype req/1. req := 'DELETE' | 'GET' | 'POST' | 'PUT'.
:- regtype res/1. res := 'OK' | 'CREATED' | 'BAD_REQUEST' | 'NOT_FOUND'.
\end{lstlisting}
and we add the following higher-order assertion to the server predicate:
\begin{lstlisting}[firstnumber=5]
:- pred server(H,Rq,Rs) : (handler(H), req(Rq)) => res(Rs).
\end{lstlisting}
{\itshape\hiord} detects that the predicate \texttt{h/2} does not
definitely conform to \ciaoinline{handler} (\texttt{h/2} $\aconfover$
\ciaoinline{handler}) due to one its clauses:
\begin{lstlisting}[firstnumber=6]
h('PUT', Rs) :- ..., Rs = 'BAD_REQ'. % Bug, should be 'BAD_REQUEST'
\end{lstlisting}
\subsection{Dutch national flag}
This problem involves sorting a list of red, white, or blue elements,
such that elements of the same color are grouped together in a
specified order (typically red, then white, then blue).
However, we want to generalize the solution by allowing the user to
provide a comparator that, given two elements, yields their
comparison.
We first define regular types to represent colored elements and
the result of their comparison:

\noindent\begin{minipage}{.48\textwidth}
\begin{lstlisting}
:- regtype rwb/1. rwb := r | w | b.
\end{lstlisting}
\end{minipage}
\hfill
\begin{minipage}{.48\textwidth}
\begin{lstlisting}[firstnumber=2]
:- regtype lge/1. lge := < | > | = .
\end{lstlisting}
\end{minipage}

\noindent Next, we define a \ciaoinline{dutch_cmp} predicate property
describing comparators between \ciaoinline{rwb} elements; and provide
a higher-order assertion to the \ciaoinline{dutch_flag/3} higher-order
predicate:
\begin{lstlisting}[firstnumber=3]
dutch_cmp := { :- pred _(X,R,Y) : (rwb(X), rwb(Y)) => lge(R). }.
:- pred dutch_flag(C,Xs,Ys) : (dutch_cmp(C), list(rwb,Xs)) => list(rwb,Ys).
\end{lstlisting}
Assume that we are given the implementation of
\ciaoinline{dutch_flag/3} and we need to provide a comparator
\texttt{cmp/3} which conforms to \ciaoinline{dutch_cmp}.
Consider a first implementation attempt:

\noindent\begin{minipage}{.6\textwidth}
\begin{lstlisting}[firstnumber=6]
cmp(red,=,  red). cmp(white,=,white). cmp(blue,=, blue).
cmp(red,<,white). cmp(white,>,  red). cmp(blue,>,  red).
cmp(red,<, blue). cmp(white,<, blue). cmp(blue,>,white).
\end{lstlisting}
\end{minipage}
\hspace*{3.2em}
\begin{minipage}{.2\textwidth}
\begin{lstlisting}
:- regtype rt1/1.
rt1 := red | blue
     | white.
\end{lstlisting}
\end{minipage}

\noindent When determining its conformance to \ciaoinline{dutch_cmp},
{\itshape\hiord} finds that \texttt{cmp/3} $\naconf$
\ciaoinline{dutch_cmp}, since \ciaopp infers the regular type
\ciaoinline{rt1} for the elements to compare, and \ciaoinline{rt1}
$\sqcap$ \ciaoinline{rwb} $=\bot$ in \textsf{eterms}.
We proceed by correcting it, but we accidentally mistype some of the
\texttt{r} elements for \texttt{o} elements in
lines~\ref{lst:cmp:line1} and~\ref{lst:cmp:line2}; and \ciaopp infers
the following assertion and regular type:

\noindent\begin{minipage}{.65\textwidth}
\begin{lstlisting}[firstnumber=6]
:- pred cmp(X,R,Y) : (rt2(X), rwb(Y)) => lge(R).
cmp(o,=,r).$\label{lst:cmp:line1}$ cmp(w,=,w). cmp(b,=,b).
cmp(o,<,w).$\label{lst:cmp:line2}$ cmp(w,>,r). cmp(b,>,r).
cmp(r,<,b). cmp(w,<,b). cmp(b,>,w).
\end{lstlisting}
\end{minipage}
\hspace*{1.205em}
\begin{minipage}{.3\textwidth}
\begin{lstlisting}
:- regtype rt2/1.
rt2 := r | b
     | w | o.
\end{lstlisting}
  \vspace*{.9em}
\end{minipage}

\noindent However, {\itshape\hiord} now reports that \texttt{cmp/3}
$\aconfover$ \ciaoinline{dutch_cmp}, since it would not raise a
run-time check error when called with an \texttt{o} on its first
argument.
Formally, \ciaoinline{(rt2(X), rwb(Y))} $\sqsupseteq$
\ciaoinline{(rwb(X), rwb(Y))} in \textsf{eterms}.
In an attempt at improving the precision of the ordering, we refine
\texttt{cmp/3} to yield more informative results on the order relation
between elements:

\noindent\begin{minipage}{.69\textwidth}
\begin{lstlisting}[firstnumber=6]
:- pred cmp(X,R,Y) : (rwb(X), rwb(Y)) => lgLGe(R).
cmp(r, =,r). cmp(w,=,w). cmp(b, =,b).
cmp(r, <,w). cmp(w,>,r). cmp(b,>>,r).
cmp(r,<<,b). cmp(w,<,b). cmp(b, >,w).
\end{lstlisting}
\end{minipage}
%
% \hspace*{em}
%
\begin{minipage}{.3\textwidth}
\begin{lstlisting}[firstnumber=10]
:- regtype lgLGe/1.
lgLGe :=  < |  >
       | << | >>
       |  = .
\end{lstlisting}
\end{minipage}

\noindent In particular, we introduce \ciaoinline{<<} and
\ciaoinline{>>} to reflect that \texttt{X} is ``much lower'' than
\texttt{Y}, and vice-versa; and define a new regular type and
assertion.
However, {\itshape\hiord} still reports that \texttt{cmp/3}
$\aconfover$ \ciaoinline{dutch_cmp}, since \texttt{cmp/3} may yield
comparison results that are not reflected in \ciaoinline{lge}.
Formally \ciaoinline{lgLGe(R)} $\sqsupseteq$ \ciaoinline{lge(R)} in
\textsf{eterms}.
Finally, we develop the following comparator:
\begin{lstlisting}[firstnumber=6]
:- pred cmp(X, R, Y) : (rwb(X), rwb(Y)) => lge(R).
cmp(r,=,r). cmp(w,=,w). cmp(b,=,b).
cmp(r,<,w). cmp(w,>,r). cmp(b,>,r).
cmp(r,<,b). cmp(w,<,b). cmp(b,>,w).
\end{lstlisting}
And {\itshape\hiord} proves that \texttt{cmp/3} $\aconf$
\ciaoinline{dutch_cmp}, since it behaves exactly as expected.
\section{Conclusions}
\label{sec:conc}
We have presented {\itshape\hiord}, a novel approach for the
compile-time verification of higher-order (C)LP programs with
higher-order assertions.
We started by refining both the syntax and semantics of predicate
properties.
Then, we introduced an abstract criterion to determine whether a
predicate conforms with a predicate property at compile time.
We also motivated and explained a \emph{wrapper}-based technique for
``casting'' predicate usage in a program analysis-friendly manner that
enhances and complements the proposed abstract criterion.
We then proposed a technique for dealing with these properties using
an abstract interpretation-based static analyzer for programs with
first-order assertions.
Finally, we reported on a prototype implementation and studied the
effectiveness of the approach with various examples within the \ciao
system.
We believe our proposal constitutes a practical approach to closing
the existing gap in the verification at compile time of higher-order
assertions; and that it is quite general and flexible, and can be
applied, at least conceptually, to other similar gradual approaches.
\section*{Competing interests}
The authors declare none.
\bibliographystyle{tlplike}
\bibliography{extracted}
\clearpage
\appendix%
\section{Proof of~\cref{th:absconf}}
\label{app:proof:conf}
We start with some auxiliary lemmas.
\renewcommand*{\thelemma}{A.\arabic{lemma}}
\begin{lemma}
  \label{app:lem:con-set-op}
  Given two property formulas $F_1$ and $F_2$,
  \(
  \trivset{(F_1 \wedge F_2)}
  = \trivset{F_1} \cap \trivset{F_2}.
  \)
\end{lemma}
\emph{Proof}.\\
We proceed by direct proof.
Assume $\trivset{(F_1 \wedge F_2)} \neq \emptyset$, take any
$\theta \in \trivset{(F_1 \wedge F_2)}$.
By~\cref{def:trivial-suc-set}, $\trivSucc{(F_1 \wedge F_2)}$,
and from~\cref{def:succeeds-triv}, $\trivSucc{F_1} \wedge \trivSucc{F_1}$.
Hence, by~\cref{def:trivial-suc-set},
$\theta \in \trivset{F_1} \wedge \theta \in \trivset{F_2}$, and from
the properties of set intersection ($\cap$), we conclude
$\theta \in (\trivset{F_1} \cap \trivset{F_2})$.
Assume now $\trivset{(F_1 \wedge F_2)} = \emptyset$, then
$\forall \theta \in \trivset{F_1} \quantSep \theta \notin
\trivset{F_2}$; hence
$\trivset{F_1} \cap \trivset{F_2} = \emptyset$.
\envend
\begin{lemma}
  \label{app:lem:cover}
  Given the calls assertion condition $\acCall{\p}{\pre} \in \A$, and
  the anonymous calls assertion condition
  $\acCall{\anonyHead}{\anonyPre} \in \Pi$ associated to $\Pi$:
  \[
    \abstrivsupset{\anonyPre}{} \sqsubseteq \abstrivsubset{\pre}{}
    \implies
    p \textit{ can be covered with } \Pi.
  \]
\end{lemma}
\emph{Proof}.\\
The proof proceeds by assuming
$\abstrivsupset{\anonyPre}{} \sqsubseteq \abstrivsubset{\pre}{}$
and showing $\trivset{\anonyPre} \subseteq \trivset{\pre}$.
\begin{flalign*}
  &\; \abstrivsupset{\anonyPre}{} \sqsubseteq \abstrivsubset{\pre}{}
  &\\
  \iff &\; \gamma(\abstrivsupset{\anonyPre}{}) \subseteq \gamma(\abstrivsubset{\pre}{})
  & \Lbag \text{Abstract operator ($\sqsubseteq$)} \Rbag\\
  \iff &\; \trivset{\anonyPre} \subseteq
           \gamma(\abstrivsupset{\anonyPre}{}) \subseteq
           \gamma(\abstrivsubset{\pre}{}) \subseteq
           \trivset{\pre}
  & \Lbag \text{Defs.~\ref{def:abstract-trivial-sub-set}
    and~\ref{def:abstract-trivial-sup-set}} \Rbag\\
  \iff &\; \trivset{\anonyPre} \subseteq \trivset{\pre}
  & \Lbag \text{Partial order ($\subseteq$)} \Rbag
\end{flalign*}
$\Box$\\[2mm]\noindent
\renewcommand*{\thetheorem}{\arabic{section}.\arabic{theorem}}
\setcounter{section}{4}
\setcounter{theorem}{0}
\renewcommand*{\theequation}{A.\arabic{equation}}
We now proceed to the proof of~\cref{th:absconf}.
\begin{theorem}
  Let $p$ be a predicate, $\Pi$ be a predicate property:
  $p \aconf \Pi \Rightarrow p \conf \Pi$, and
  $p \naconf \Pi \Rightarrow p \nconf \Pi$.
\end{theorem}
\smallskip\emph{Proof.}\\
Let $\prog$ be a program, $p$ be a predicate s.t.~$p \in \prog$, and
$\Pi$ be a predicate property.
We will prove each case separately.
\begin{itemize}
\item We proceed by contradiction.
  Assume that $p$ abstractly conforms to $\Pi$:
  \begin{equation}
    p \aconf \Pi\label{proof:aconf}
  \end{equation}
  and that $p$ does not conform to $\Pi$:
  \begin{equation}
    p \nconf \Pi\label{proof:nconf}
  \end{equation}
  First, by~\cref{proof:aconf} and~\cref{app:lem:cover},
  we know that $p$ can be \emph{covered} with $\Pi$, hence
  \begin{equation}
    \trivset{\anonyPre} \subseteq \trivset{\pre}\label{proof:cover}
  \end{equation}
  From~\cref{proof:nconf} we know that there exists a query
  $\cquery_{p} = (\p,\theta_0)$ to $p$ s.t.~$\Pi$ is \emph{not} redundant
  for $p$.
  Thus, by negating the redundance condition
  from~\cref{def:redundance}, we know that:
  \(
  \exists D' \in \derivsA[\A']{\prog}{\cquery_{p}} \quantSep D'_{[-1]} =
  \exstate{G'}{\theta}{\{\ell'\}},
  \)
  and
  \(
  \exists D \in \derivsA{\prog}{\cquery_{p}} \quantSep\allowbreak\mathcal{U}(D) =
  \mathcal{U}(D'),
  \)
  s.t.\
  \(
  \redstarA{D_{[-1]}}{\exstate{G}{\theta}{\emptyset}}
  \)
  through any derivation that reduces only $\mathsf{check}$
  literals; where $\A'$ is defined as in~\cref{def:conformance}.
  We also know that both $\derivsA[\A']{\prog}{\cquery_{p}}$ and
  $\derivsA{\prog}{\cquery_{p}}$ share the same initial state
  $S = \exstate{\p}{\theta_0}{\emptyset}$ given that they are deriving
  the same query $\cquery_{p}$ to $p$.
  We now consider two cases: the error comes from a $\mathsf{calls}$
  or $\mathsf{success}$ assertion condition in $\A'$.

  \medskip
  \underline{\textsf{\bfseries Calls}}: The label $\ell'$ identifies a
  $\mathsf{calls}$ assertion condition $C'$ in $\A'$ -- of the
  form $\acCall{\p}{\pre \wedge \anonyPre}$ (from the definition of
  $\A'$) -- which fails to be checked for $\theta$; and the
  $\mathsf{calls}$ assertion condition $C$ in $\A$ -- of the form
  $\acCall{\p}{\pre}$ -- is checked for $\theta$.
  On the one hand:
  \begin{flalign*}
    &\; C' \text{ fails to be checked for } \theta
    & \\
    \iff &\; \notTrivSucc{\pre \wedge \anonyPre}
    & \Lbag \text{Semantics with assertions} \Rbag\\
    \iff &\; \theta \notin \trivset{(\pre \wedge \anonyPre)}
    & \Lbag \text{Def.~\ref{def:trivial-suc-set}} \Rbag\\
    \iff &\; \theta \notin \trivset{\pre} \cap \trivset{\anonyPre}
    & \Lbag \text{Lem.~\ref{app:lem:con-set-op}} \Rbag\\
    \iff &\; \theta \notin \trivset{\anonyPre}
    & \Lbag \text{Eq.~(\ref{proof:cover})} \Rbag
  \end{flalign*}
  On the other hand:
  \begin{flalign*}
    &\; C \text{ is checked for } \theta
    & \\
    \iff &\; \trivSucc{\pre}
    & \Lbag \text{Semantics with assertions} \Rbag\\
    \iff &\; \theta \in \trivset{\pre}
    & \Lbag \text{Def.~\ref{def:trivial-suc-set}} \Rbag
  \end{flalign*}
  Since $p \aconf \Pi$ (by~\cref{proof:aconf}), we know
  from~\cref{def:conf-call} that:
  \begin{flalign*}
    &\; \abstrivsupset{\pre}{} \sqsubseteq \abstrivsubset{\anonyPre}{}
    & \\
    \iff &\; \gamma(\abstrivsupset{\pre}{}) \subseteq \gamma(\abstrivsubset{\anonyPre}{})
    & \Lbag \text{Abstract operator ($\sqsubseteq$)} \Rbag\\
    \iff &\; \trivset{\pre} \subseteq
           \gamma(\abstrivsupset{\pre}{}) \subseteq
           \gamma(\abstrivsubset{\anonyPre}{}) \subseteq
           \trivset{\anonyPre}
    & \Lbag \text{Defs.~\ref{def:abstract-trivial-sub-set}
      and~\ref{def:abstract-trivial-sup-set}} \Rbag
  \end{flalign*}
  Since $\theta \in \trivset{\pre}$ and $\subseteq$ is a partial order,
  it implies that $\theta \in \trivset{\anonyPre}$, contradicting
  $\theta \notin \trivset{\anonyPre}$.  \Lightning

  \medskip

  \underline{\textsf{\bfseries Success}}: The label $\ell'$
  identifies a $\mathsf{success}$ assertion condition $C'$ in
  $\A'$ -- of the form $\acSucc{\p}{\anonyPre}{\anonyPost}$ -- which
  fails to be checked for $\theta$; and all $\mathsf{success}$
  assertion conditions $C_i$ in $\A$ -- of the form
  $\acSucc{\p}{\pre_i}{\post_i}$ with $i \in 1..n$ -- is checked for
  $\theta$.
  On the one hand:
  \begin{flalign*}
    &\; C' \text{ fails to be checked for } \theta
    & \\
    \iff &\; (\trivSucc[\theta_0]{\anonyPre})
           \wedge
           (\notTrivSucc{\anonyPost})
    & \Lbag \text{Semantics with assertions} \Rbag\\
    \iff &\; (\theta_0 \in \trivset{\anonyPre})
           \wedge
           (\theta \notin \trivset{\anonyPost})
    & \Lbag \text{Def.~\ref{def:trivial-suc-set}} \Rbag
  \end{flalign*}
  Since $p \aconf \Pi$ (by~\cref{proof:aconf}), we know
  from~\cref{def:conf-succ} that there exists a proper subset of
  $\mathsf{success}$ assertion conditions $S \subset \A$ s.t.:
  \begin{equation}
    (\pre_{\sqcup}^{\sharp-} =
    \sqcup\setcomp{\abstrivsubset{\pre}{}}{\acSucc{\p}{\pre}{\_} \in S})
    \sqsupseteq \abstrivsupset{\anonyPre}{}\label{proof:chain1}
  \end{equation}
  and
  \begin{equation}
    (\post_{\sqcup}^{\sharp+} =
    \sqcup\setcomp{\abstrivsupset{\post}{}}{\acSucc{\p}{\_}{\post} \in S})
    \sqsubseteq \abstrivsubset{\anonyPost}{}\label{proof:chain2}
  \end{equation}
  From~\cref{proof:chain1,proof:chain2,def:abstract-trivial-sub-set,def:abstract-trivial-sup-set},
  we obtain the following relations:
  \begin{equation}
    \cup\setcomp{\trivset{\pre}}{\acSucc{\p}{\pre}{\_} \in S}
    \supseteq
    \gamma(\pre_{\sqcup}^{\sharp-})
    \supseteq
    \gamma(\abstrivsupset{\anonyPre}{})
    \supseteq
    \trivset{\anonyPre}\label{proof:chain3}
  \end{equation}
  and
  \begin{equation}
    \cup\setcomp{\trivset{\post}}{\acSucc{\p}{\_}{\post} \in S}
    \subseteq
    \gamma(\post_{\sqcup}^{\sharp+})
    \subseteq
    \gamma(\abstrivsubset{\anonyPost}{})
    \subseteq
    \trivset{\anonyPost}\label{proof:chain4}
  \end{equation}
  Finally, since $\theta_0 \in \trivset{\anonyPre}$:
  \begin{flalign*}
    &\; \theta_0 \in \trivset{\anonyPre}
    & \\
    \iff &\; \theta_0 \in \cup\setcomp{\trivset{\pre}}{\acSucc{\p}{\pre}{\_} \in S}
    & \Lbag \text{Eq.~(\ref{proof:chain3})} \Rbag\\
    \iff &\; \exists (C = \acSucc{\p}{\pre}{\post}) \in S \quantSep \theta_0 \in \trivset{\pre}
    & \Lbag \text{Set union ($\cup$)} \Rbag\\
    \iff &\; \trivSucc{\post}
    & \Lbag C \text{ checked for } \theta \text{, hypothesis} \Rbag\\
    \iff &\; \theta \in \trivset{\post}
    & \Lbag \text{Def.~\ref{def:trivial-suc-set}} \Rbag\\
    \iff &\; \theta \in \cup\setcomp{\trivset{\post}}{\acSucc{\p}{\_}{\post} \in S}
    & \Lbag \text{Set union ($\cup$)} \Rbag\\
    \iff &\; \theta \in \trivset{\anonyPost}
    & \Lbag \text{Eq.~(\ref{proof:chain4})} \Rbag
  \end{flalign*}
  which leads to a contradiction with our initial hypothesis
  $\theta \notin \trivset{\anonyPost}$.  \Lightning\\\smallskip
  
\item We proceed by direct proof.
  Since $p$ does not abstractly conform to $\Pi$:
  \begin{equation}
    p \naconf \Pi\label{proof:naconf}
  \end{equation}
  we consider two cases: $p$ does not abstractly conform to $\Pi$'s
  anonymous $\mathsf{calls}$ assertion condition, or to some anonymous
  $\mathsf{success}$ assertion condition of $\Pi$.

  \medskip
  \underline{\textsf{\bfseries Calls}}: Let
  $(\anonyACond = \acCall{\anonyHead}{\anonyPre}) \in \Pi$, we know
  that $p \naconf \anonyACond$, then:
  \begin{flalign*}
    &\; p \naconf \anonyACond
    & \\
    \iff &\; \abstrivsupset{\pre}{} \sqcap \abstrivsupset{\anonyPre}{} = \bot
    & \Lbag \text{Def.~\ref{def:conf-call}} \Rbag\\
    \iff &\; \trivset{\pre} \cap \trivset{\anonyPre} = \emptyset
    & \Lbag \text{Def.~\ref{def:abstract-trivial-sup-set}} \Rbag\\
    \iff &\; \forall \theta \in \trivset{\anonyPre} \quantSep \theta \notin \trivset{\pre}
    & \Lbag \text{Set intersection ($\cap$)} \Rbag\\
    \implies &\; \trivset{\anonyPre} \not\subset \trivset{\pre}
    & \Lbag \text{Proper subset ($\subset$)} \Rbag\\
    \implies &\; p \textit{ cannot be covered with } \Pi
    & \Lbag \text{Def.~\ref{def:covered}} \Rbag\\
    \implies &\; p \nconf \Pi
    & \Lbag \text{Defs.~\ref{def:redundance} and~\ref{def:conformance}} \Rbag
  \end{flalign*}

  \medskip
  \underline{\textsf{\bfseries Success}}: Let
  $(\anonyACond = \acSucc{\anonyHead}{\anonyPre}{\anonyPost}) \in
  \Pi$, since $p \naconf \anonyACond$, we know that there exists a $\mathsf{success}$
  assertion condition $(C = \acSucc{\p}{\pre}{\post}) \in \A$ s.t.~the following
  hold:
  \begin{align}
    \abstrivsupset{\pre}{} &\sqsubseteq \abstrivsubset{\anonyPre}{}\label{proof:chain5}\\
    \abstrivsupset{\post}{} &\sqcap \abstrivsupset{\anonyPost}{} = \bot\label{proof:chain6}\\
    \exists \theta_0 \in \trivset{\pre} \quantSep
                           & \mathcal{S}_{\A}(\p,\theta_0,\prog,\gamma(\aqueries_{p})) \neq \emptyset\label{proof:chain7}
  \end{align}
  where $\gamma(\aqueries_{p})$ is a \emph{superset} of the set of valid
  queries to $p$ described by $p$'s $\mathsf{calls}$ assertion
  condition in $\A$.
  First,
  by~\cref{def:abstract-trivial-sub-set,def:abstract-trivial-sup-set},
  from~\cref{proof:chain5} and~\cref{proof:chain6} we obtain:
  \begin{align}
    \trivset{\pre} &\subseteq \trivset{\anonyPre}\label{proof:chain8}\\
    \trivset{\post} &\cap \trivset{\anonyPost} = \emptyset\label{proof:chain9}
  \end{align}
  Next, from~\cref{proof:chain7} and \cref{proof:chain8}, it follows
  that $\theta_0 \in \trivset{\pre}$ and
  $\theta_0 \in \trivset{\anonyPre}$.
  Also, from~\cref{proof:chain7} and the definition of the
  \emph{success context}, we know that
  \begin{equation*}
    \theta \in
    \mathcal{S}_{\A}(\p,\theta_0,\prog,\gamma(\aqueries_{p}))
  \end{equation*}
  hence $\theta \in \trivset{\post}$.
  However, by~\cref{proof:chain9} we know that
  $\theta \notin \trivset{\anonyPost}$.
  By~\cref{def:trivial-suc-set}, this means that, on the one hand:
  \begin{equation}
    \trivSucc[\theta_0]{\pre} \wedge \trivSucc{\post}\label{proof:chain10}
  \end{equation}
  And in the other hand:
  \begin{equation}
    \trivSucc[\theta_0]{\anonyPre} \wedge \notTrivSucc{\anonyPost}\label{proof:chain11}
  \end{equation}
  The above conditions~\cref{proof:chain10} and~\cref{proof:chain11},
  together with the definition of the operational semantics with
  assertions, implies that there exists a query
  $\cquery_{p} = (\p,\theta_0)$ to $p$ s.t.~$\Pi$ is \emph{not}
  redundant for $p$.
  And, by the definition of \emph{non conformance}
  (\cref{def:conformance}), this implies that $p$ \emph{does not}
  conform to $\Pi$.  \hfill$\Box$
\end{itemize}
\end{document}